\documentclass[aps,prl,twocolumn,letterpaper,superscriptaddress]{revtex4}
\usepackage[colorlinks,linkcolor=blue,anchorcolor=blue, citecolor=blue,urlcolor=blue,]{hyperref}
\usepackage{graphicx}
\usepackage{CJK}
\usepackage{verbatim}
\usepackage{physics}
\usepackage{color}
\usepackage{mathptmx}
\usepackage{amssymb}
\usepackage{amsmath}

\begin{document}

\begin{CJK*}{UTF8}{bsmi}
\title{Geometric frustration produces long-sought Bose metal phase of quantum matter}
\author{Anthony Hegg}
\affiliation{Tsung-Dao Lee Institute, Shanghai Jiao Tong University, Pudong, Shanghai 200240, China}
\author{Jinning Hou (\CJKfamily{gbsn}侯晋宁)}
\affiliation{Tsung-Dao Lee Institute, Shanghai Jiao Tong University, Pudong, Shanghai 200240, China}
\affiliation{School of Physics and Astronomy, Shanghai Jiao Tong University, Shanghai 200240, China}
\author{Wei Ku (\CJKfamily{bsmi}顧威)}
\altaffiliation{corresponding email: weiku@sjtu.edu.cn}
\affiliation{Tsung-Dao Lee Institute, Shanghai Jiao Tong University, Pudong, Shanghai 200240, China}
\affiliation{School of Physics and Astronomy, Shanghai Jiao Tong University, Shanghai 200240, China}
\affiliation{Ministry of Education Key Laboratory of Artificial Structures and Quantum Control, Shanghai 200240, China} 

\date{\today}

\begin{abstract}
Two of the most prominent phases of bosonic matter are the superfluid with perfect flow and the insulator with no flow. A now decades-old mystery unexpectedly arose when experimental observations indicated that bosons could organize into the formation of an entirely different intervening third phase: the Bose metal with dissipative flow. The most viable theory for such a Bose metal to date invokes the use of the extrinsic property of impurity-based disorder; however, a generic intrinsic quantum Bose metal state is still lacking. We propose a universal homogeneous theory for a Bose metal in which geometric frustration confines the essential quantum coherence to a lower dimension. The result is a gapless insulator characterized by dissipative flow that vanishes in the low-energy limit. This failed insulator exemplifies a frustration-dominated regime that is only enhanced by additional scattering sources at low energy and therefore produces a Bose metal that thrives under realistic experimental conditions.
\end{abstract}

\maketitle
\end{CJK*}

Superfluidity has fascinated many with the phenomenon of perfect (dissipationless) flow since its discovery\cite{KapitzaSF,AllenMisenerSF}. One of the few homogeneous phases of matter known to disrupt this low-temperature behavior is the Mott insulator\cite{fwgf,Greiner} characterized by a complete lack of flow. Bosons underlie both phases of matter and are well-known to exhibit other exotic and extreme properties such as Bose-Einstein condensation. For decades it had been understood that bosons can undergo a transition from the perfect flow of superfluidity directly into the Mott insulating state of no flow, exemplifying this extreme nature. In the absence of inhomogeneities this abrupt transition was understood to dominate low-energy bosonic systems in general. 

Therefore, an unexpected mystery arose over the last several decades as experiments\cite{Goldmanmetaldata,BMExpt1,BMExpt2,BMExpt4,BMExpt3,Bozovich2e} have continued to find evidence that seemingly disparate bosonic systems exhibit more mundane dissipative transport. Surprisingly, the prevailing theories could not account for this behavior whatsoever. In the intervening decades, several theories were developed to fill this unexpected gap, but despite these efforts there remains no consensus even for a qualitative account of these observations. Identifying a universal mechanism that allows for a stable phase of dissipative bosonic transport at low-temperature remains a holy grail of condensed matter physics and ultra-cold atom research. 

The extreme nature of bosons has guided theoretical developments toward several specialized directions as opposed to a universal qualitative understanding of this metallic phase. Early research focused on superconducting grain models\cite{ppdd33,ppdd34,ppdd35,ppdd36}, but apparent metallic behavior was shown to be unstable at finite temperature\cite{ppdd38} and a large class of models were further ruled out by scaling arguments\cite{ppdd37}. Another approach used exotic interactions on a lattice such as ring exchange models\cite{ppdd39,Tay}, but these are unstable toward insulating phases with the realistic introduction of weak disorder\cite{ppdd40}. A more recent approach involves the so-called moatband models\cite{GorkovRashba,Moatband1,Moatband2,Moatband3,Moatband4,Moatband5,Kun}. These models are solved in the extreme dilute limit where the particle density scales much slower than the volume, which is not the regime relevant to the experimental observations. Even more recently, a phenomenological field theory has been developed to account for metallic crossover behavior in 2D systems\cite{2DTopoIntBoseMetal}, but such methods would be inapplicable deep inside a stable phase of matter. A top contender\cite{ppdd45} applies extrinsic phase frustration to avoid the issues of the above systems but leaves the question of a universal intrinsic mechanism unanswered. 

Instead of attempting to tame the extreme tendencies of bosons by hand, we propose a universal picture where both extremes coexist and stabilize metallic behavior. Consider an emergent system of freely flowing bosons composed of lower dimensional subsystems. If flow between subsystems vanishes at low energy due to perfect phase interference, then each subsystem is effectively disconnected. We find a gapless insulating state at zero temperature protected by complete frustration of the quantum phases that becomes metallic upon introduction of finite temperature or weak disorder. This failed insulator is mediated by a dimensional crossover between freely moving bosons and disconnected subsystems, providing a universal origin for stable metallic behavior, so we expect this type of Bose metal to be ubiquitous in nature. 

We consider a relatively simple implementation of this Bose metal by stacking 2D checkerboard lattice $xy$-layers along the $z$-axis to form a 3D lattice. Alternatively, this system can be viewed as two overlapping sets of stacked vertical $xz$- and $yz$- slabs. We study the regime in which the hopping between such vertical slabs is weaker than that within each slab. In the low-energy limit these vertical slabs become disconnected. This model has been suggested\cite{prepairmodelweiku} to represent the underdoped cuprates beyond the superconducting dome, a class of materials suspected to exhibit Bose metal behavior through the formation of an emergent Bose liquid (EBL)\cite{prepairmodelweiku,yildirimWeiku,xander1,laurence1,xinlei1}.  Even if we drive each slab into superfluidity by introducing a small repulsive local interaction, transport between slabs is still suppressed at low energy. Although proving that a superfluid phase exists can be somewhat subtle (see e.g. \cite{leggettQL} for a discussion on this issue), proving that it does not exist is far less stringent. The lack of transport between effectively independent slabs at low energy forbids superflow, and upon introduction of temperature or disorder we find a Bose metal as a failed insulator.

\section*{Results and Discussion}

\begin{figure}[ht!]
\centering
\includegraphics[scale=0.17]{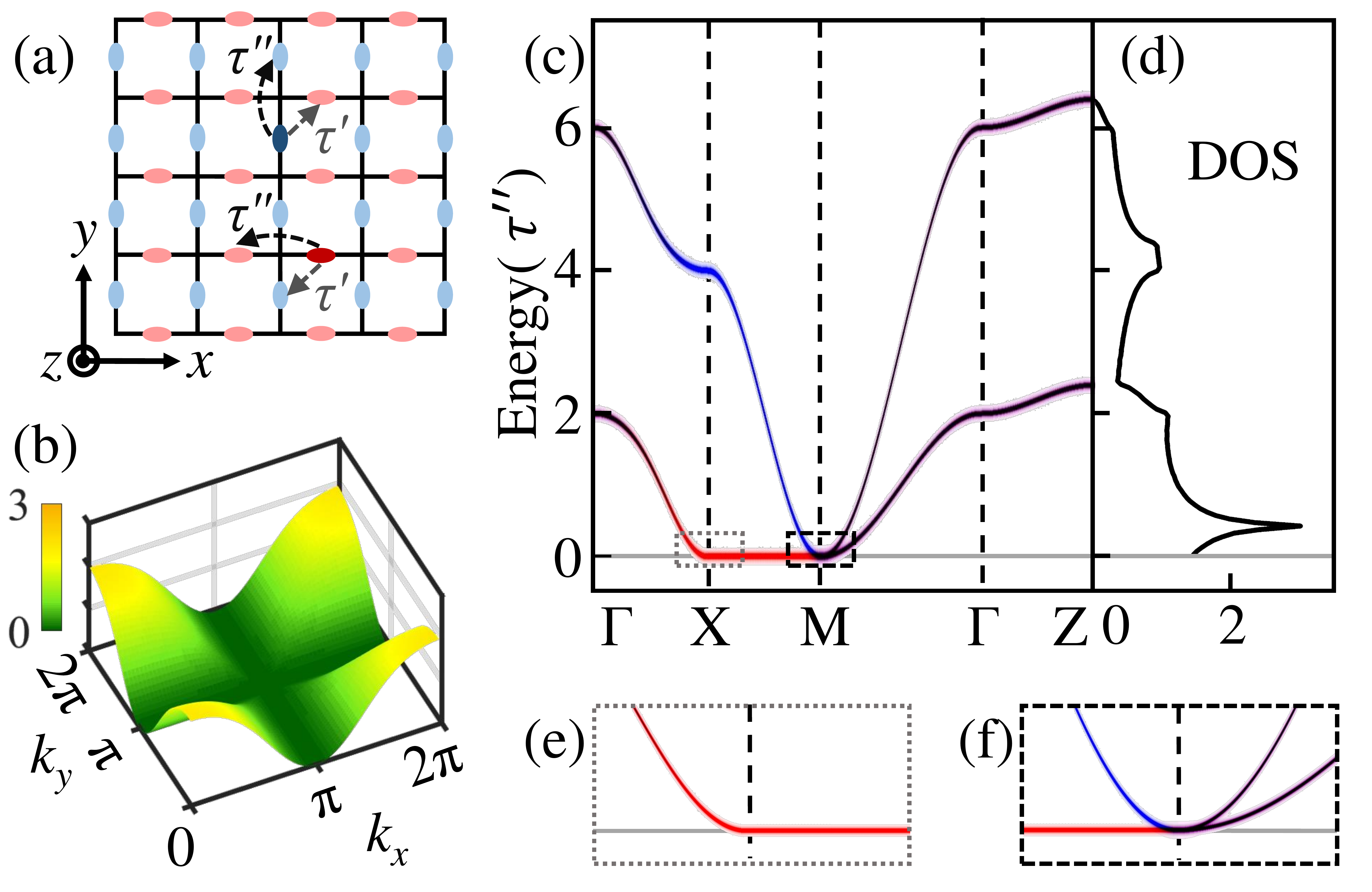}
\caption{The checkerboard lattice. (a) There are two sublattices, denoted red and blue for horizonal and vertical site orientation respectively, with nearest neighbor hopping $\tau'>0$ between sublattices and next nearest neighbor hopping $\tau''$ following site orientation. (b) A symptom of the frustrated regime ($\tau''>\tau'$) is the appearance of a line degeneracy in the band minimum denoted in green. (c) The corresponding dispersion with orbital weight given in red and blue as in (a) with $\Gamma$, X, M and Z representing ($0,0,0$), ($\pi,0,0$), ($\pi,\pi$,0) and ($0,0,\pi$) respectively. Note the clear separation of orbitals along the line degeneracy. (d) The 3D density of states (DOS in states per 100$\tau''$ per unit cell) converge to a constant at low energy due to the emergent two dimensional nature of the excitations near the line degeneracy. (e-f) Magnification at low energy shows that the low-lying excitations are quadratic.}
\label{fig1}
\end{figure}

The intralayer structure of the 2D checkerboard lattice as well as its inherent two-band nature is shown in Fig.(\ref{fig1}). In the $xy$-layer, this lattice contains hopping to four nearest-neighbors (NN) $\tau'>0$ as well as an alternating set of two next-nearest-neighbors (NNN) $\tau''>0$. We study the case where the $z$-axis has simple nearest neighbor hopping with $\tau_{z}<0$, which we suppress below for brevity. Finally, we include on-site repulsive interactions $U>0$ resulting in an $xy$-layer Hamiltonian $\mathcal{H}$ given by

\begin{align}
\label{MainH}
\mathcal{H} = \sum_{i} \left\{ \sum_{j \in \text{NN}} \tau' a^{\dag}_{i} a_{j} + \sum_{j \in \text{NNN}} \tau'' a^{\dag}_{i} a_{j} + U a^{\dag}_{i} a^{\dag}_{i} a_{i} a_{i} \right\}
\end{align}
where $a^{\dag}_{i}$ and $a_{j}$ are the bosonic creation and annihilation operators at sites $i$ and $j$ respectively. 

We study the frustrated $\tau''>\tau'$ regime of this model in the low-temperature $T \rightarrow 0$ limit. The non-interacting ($U=0$) dispersion for the two bands is given by

\begin{align}
\label{band}
&\epsilon_{\pm,\mathbf{k}} = 2\tau_{z}(\text{cos}k_z-1) + \tau''(\text{cos}k_x + \text{cos}k_y + 2) \\
\nonumber
&\pm \sqrt{4\tau'^2(1 + \text{cos}k_x)(1 + \text{cos}k_y) + \tau''^2(\text{cos}k_x - \text{cos}k_y)^2}.
\end{align}
This differs from the unfrustrated regime $\tau'>\tau''$, where there is a single lowest energy momentum state. In that case, the bosons condense into that state at low temperature and a $d$-wave superfluid obtains\cite{yildirimWeiku}. Here, as we discuss below and prove in the supplementary material, throughout the frustrated regime denoted by $\tau''>\tau'>0$, inter-slab transport via $\tau'$ is completely disabled in the low-energy limit and superflow does not occur between slabs.

\subsection*{Effectively Independent Slabs due to Frustration}

\begin{figure}[ht!]
\centering
\includegraphics[scale=0.4]{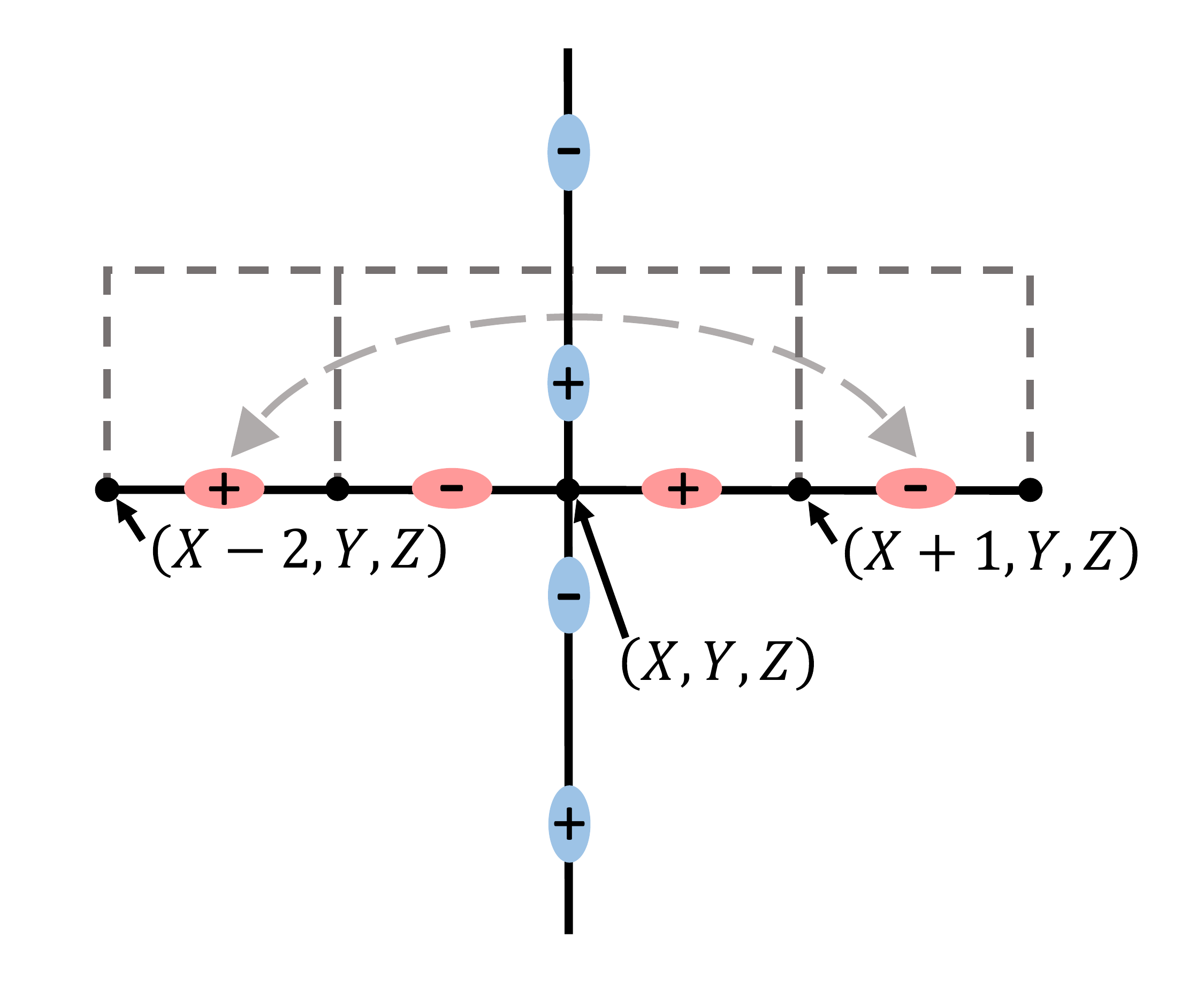}
\caption{Ground state phase structure of one $xz$-slab (red) and one $yz$-slab (blue) centered at the point where they cross within the primitive cell $(X,Y,Z)$ outlined in gray. The signs at each orbital are the phases relative to their neighbors within the same slab. Each slab hosts alternating phases along $\tau''>0$. As a result, this illustration clearly shows that the parity of the wavefunction in a single slab at the point where these two lines cross is odd along a given line, and even otherwise.}
\label{intersect}
\end{figure}

Geometrical frustration in \emph{any} lattice creates a set of states that are effectively independent from one another. Consider a special one-body state as an example, illustrated in red in Fig.(\ref{intersect}), defined in a single slab with alternating phase differences along the slab. This particular state has unusually high symmetry such that, at every point where two slabs intersect, this state has odd parity in the $x$-direction and even parity otherwise. Since parity is a good symmetry of the system, a particle in this state cannot propagate to any intersecting slab. However, this is the only path in the Hamiltonian (via $\tau'$) that allows the particle to leave the slab, so this particle is effectively localized to a lower dimension. Since the many-body dressing of a particle must respect its underlying symmetry, the confinement this state's propagation to a single slab will persist in the interacting system despite its additional quantum fluctuations in the vicinity of the slab. 

A superficial consequence of the decoupling between these states is the occurrence of a line degeneracy in momentum space \textit{along} the direction of the decoupling in the band structure. The fact that such degeneracy is present for any arbitrary choice of parameters in the Hamiltonian (cf. see Fig.(S3) in the supplementary material) reflects the profound symmetry-based origin of this decoupling. This geometrical frustration induced decoupling distinguishes the line degeneracy in this case from those in existing literature (cf. \cite{tsvelik}), since subdimensional decoupling is a much more stringent condition than just having a degenerate line in momentum space. 

Under the right conditions, when the many-body ground state is formed from the dressed version of the above confined states, one would expect physics such as the Bose metal studied here. In our checkerboard lattice example, we expect this to be the case when $\tau''>\tau'>0$ and $U$ not too big. This is clear from Fig.(\ref{fig1}), which shows that the line degeneracy, corresponding to these states, has the lowest energy under this condition at $U=0$. As long as the interaction is not so overwhelmingly strong that it drives the system into a completely different state (e.g. a Mott insulator), we expect it to simply dress the confined particles within the many-body ground state. In summary, due to geometric frustration, each particle in the many-body ground state is still unable to propagate between slabs. In other words, the slabs are effectively independent at low energy.

\subsection*{Stability}

Since the low-energy transport is dominated by the physics of each 2D slab, we can study their characteristics in order to estimate the thermal stability of this system. To facilitate this understanding we recall (see e.g. \cite{HFMeng}) that the low-energy physics of a 2D superfluid is dominated by phase-mode gapless excitations with a linear spectrum. Correspondingly, the one-body density of states (DOS) is linear at low-energy. A DOS that vanishes linearly at low energy immediately implies that the bosonic system is stable at low-temperature. Such a suppressed density of states at low energy indicates a diminishing channel for fluctuations at low temperature and therefore establishes the thermal stability of our system. 

Similarly, it is easy to verify the stability of this system against disorder. Interestingly, introduction of a single impurity breaks parity locally and ruins the perfect interference in $\tau'$. In the regime where disorder is weaker than the scale of the phase stiffness of each slab defined by $U$, such disorder will not challenge the rigidity of each slab and can only establish a small coupling between extended states in the slab. Since the effect of an impurity and the lack of coherence due to geometrical frustration are both one-body effects, it is sufficient to analyze the energy scale of the induced coherence by analyzing the scaling in a one-body Hilbert space formed by the $2L$ special states defined above. As outlined in the supplementary material, the coupling between extended states of different slabs along their lines of intersection then scales as $\mathcal{O}(1/L\sqrt{L})$ assuming that the number of impurities scale with the system size. Therefore, the leading order energy scale of the coherence scales as $\mathcal{O}(1/L^3)$. There are $L^2$ intersections between pairs of slabs, so the total relevant energy correction scales as $\mathcal{O}(L^2/L^3)=\mathcal{O}(1/L)$ and thus vanishes in the large system limit. Therefore, this coupling is not thermodynamically meaningful. This analysis does not account for the incoherent nature of disorder between different intersections, so in that sense it is an upper bound estimate of the coherence generated by impurities. In summary, disorder cannot introduce meaningful 3D coherence of the one-body states that form the many-body ground state, and the slabs remain effectively independent.

\subsection*{Our Failed Insulator is a Bose Metal}

\begin{figure}[ht!]
\centering
\includegraphics[scale=0.35]{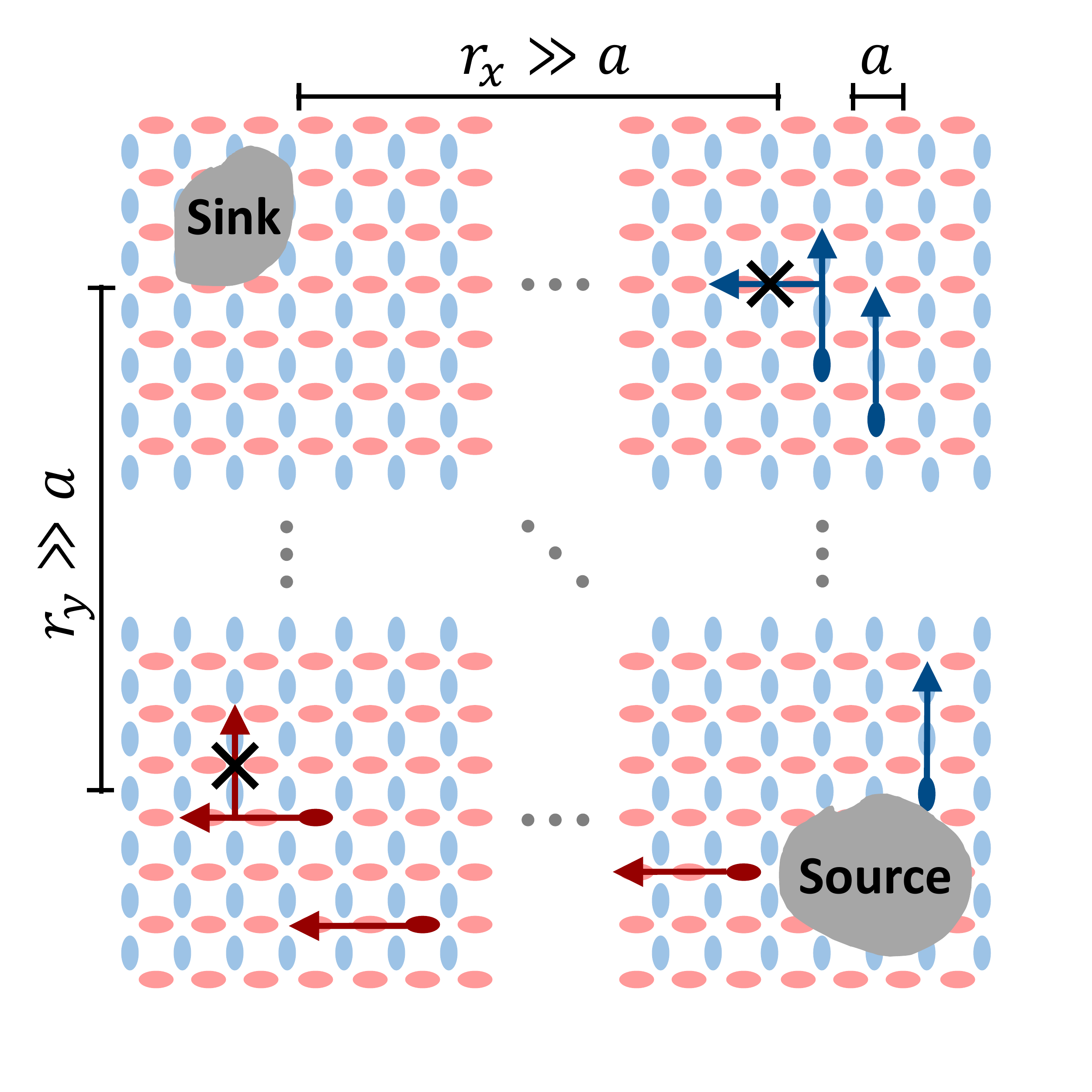}
\caption{Hypothetical experimental probe of the current with two gray contacts separated by a distance $r=\sqrt{r_{x}^2+r_{y}^2}$ much larger than the lattice spacing $a$. Particles emerge from the source and travel in straight lines along a given coherent slab. The probability for the current to flow between sublattices vanishes in the low-energy limit, so no path remains to reach the sink if either $r_{x}$ or $r_{y}$ are large enough to place the leads on distinct coherent slabs.}
\label{expt}
\end{figure}

Since each particle in the ground state cannot propagate between slabs, our system naturally does not exhibit 3D superfluidity. This is illustrated in Fig.(\ref{expt}) where, despite the superflow propagation allowed along each slab, transition between slabs is suppressed. As a result, in the low-energy limit there is no conducting path from an arbitrary source to sink, separated by distance $r=\sqrt{r_{x}^2+r_{y}^2}$ much larger than the lattice spacing $a$. Correspondingly, there is no superflow along this general path either.

On the other hand, since the ground state is formed from effectively independent slabs, naively one might expect an insulator. However, each slab hosts gapless excitations common to 2D superfluidity, so there is no energetic protection for this insulating behavior. These excitations will be modified by $\tau'$, coupling the slabs above the ground state and defeating the insulation found there. We refer to such a gapless insulator as a failed insulator. 

In this failed insulator, one could imagine finite temperature normal conductivity (in the presence of phonons and disorder for example) to be limited by the weak connection between coherent slabs. However, since there is no gap, we do not have an exponentially small activation at low temperature. Furthermore, as shown in Fig.(\ref{expt}), the path connecting sink to source can be made with as little as one $\tau'$ connection, whose leading effect is polynomial in energy. So, the temperature enhancement of the conductivity will scale faster than $e^{-\Delta/T}$ in contrast to a typical insulator with an activation scale $\Delta$. We have therefore proven that this system is neither a superfluid nor an insulator protected by an energy scale, so by definition it is a metal. Although, this metal should be distinguished from a regular metal since it has a large resistivity at low temperature even for a rather clean sample. 

Experienced readers might notice that our 2D slab could still host 2D superfluidity since it is effectively a 2D interacting bosonic subsystem. However, it is well known that 2D superfluidity is particularly susceptible to disorder in relatively low density systems (see e.g. \cite{Goldmanmetaldata,fwgf,HFMeng}). Therefore, we can expect this residual 2D superfluid response along the slab to be replaced by a low resistivity along the slab directions in the presence of weak disorder such as expected in real materials (as long as the boson density is relatively low). The resulting anisotropic transport can actually account for the puzzling observation of unexpectedly reversed anisotropy in some materials\cite{WuBozNemRuth,WuBozNemCup}.

\subsection*{Demonstration via Controlled Approximation}

To further illustrate the stable Bose metal derived above, we implement a controlled approximation in which a reference state is first established as the dominant contribution to the zero temperature many-body ground state and expand the Hamiltonian to bilinear order in fluctuations about this reference state. We construct the many-body eigenstates using the solution to the Hamiltonian in order to compute the current-current response function in the low-frequency limit. From the result we establish the conductivity and therefore the low-temperature phase of matter. 

As detailed in the supplementary material, we adopt an effective local density and phase operator formalism to represent our system. We expand the Hamiltonian in small fluctuations about a many-body reference state chosen based on the general properties of the many-body ground state identified above. In particular, we use a reference state with uniform average density and with average local phases as illustrated in Fig.(\ref{intersect}). Namely, $\pi$ phase difference across a $\tau''$ bond and overall phase freedom for each slab. 

\begin{figure}[ht!]
\centering
\includegraphics[scale=0.17]{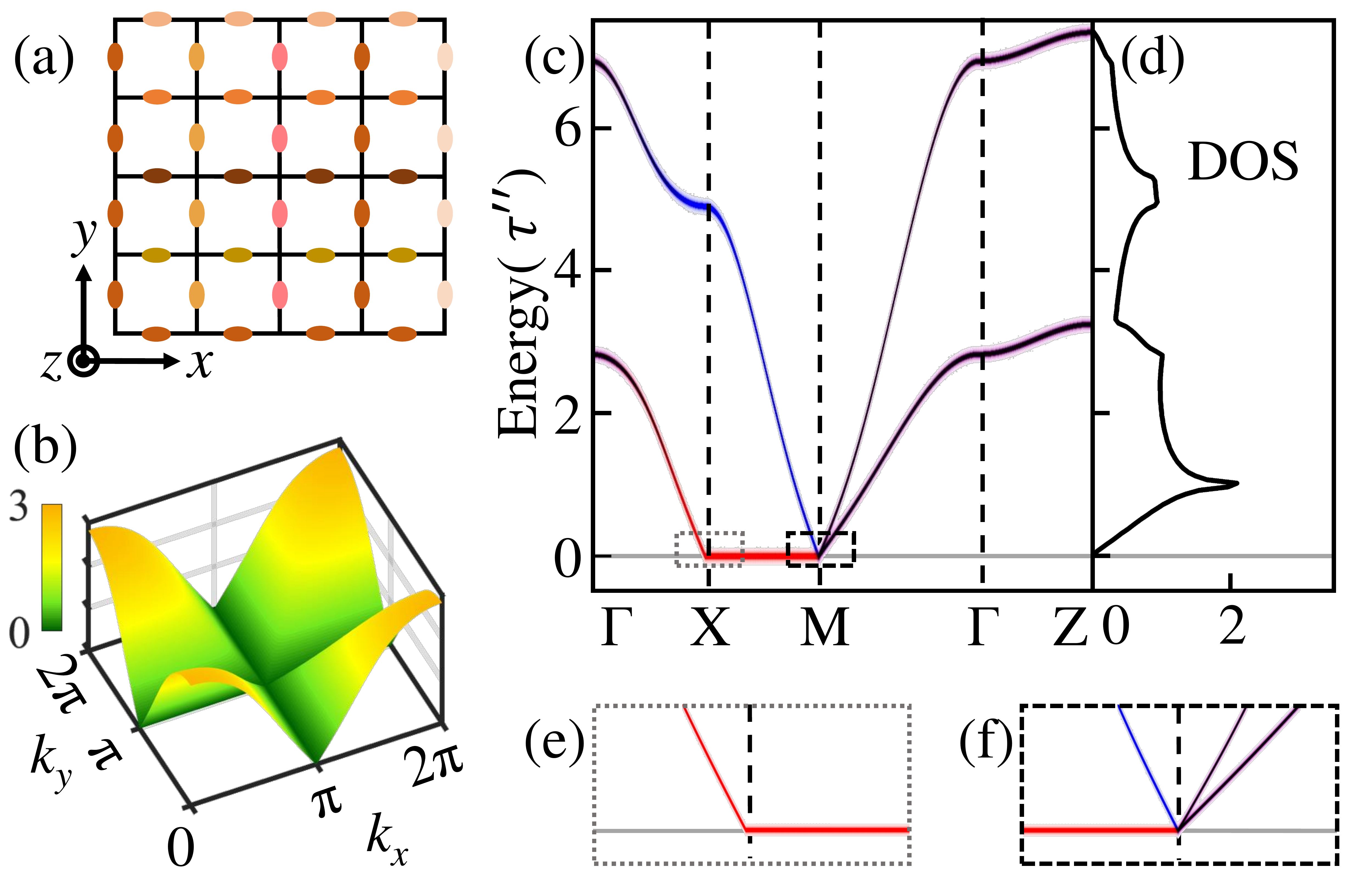}
\caption{The checkerboard lattice in the presence of local repulsive interactions $U>0$. (a) Each shade of orange represents an independent coherent slab with a randomly selected overall phase. (b-c) An example with $(\pi,\pi,0)$ periodic phase structure shows that the band structure maintains the line degeneracy and separation of orbitals from Fig.(\ref{fig1}). (d) The corresponding DOS goes to zero linearly with energy indicating that the ground state is stable at low-temperature. (e-f) The low-lying excitations are now stiffened into a linear spectrum.}
\label{fig2}
\end{figure}

As a demonstration, Fig.(\ref{fig2}) represents the solution corresponding to a coincidentally coherent phase structure
\begin{align}
\label{pipiphz}
\phi_{\alpha XYZ} &= \pi (X+Y),
\end{align}
where $\alpha \in \{ 1,2 \}$ are the two orbitals in the $(X,Y,Z)$ primitive unit cell. This choice maintains momentum as a good quantum number, and as such it provides an upper bound on the superfluid density of the system. Choosing phases that do not preserve lattice translational symmetry can only introduce additional interference and weaken any physics that relies on long-range coherence. 

All of the key features of our Bose metal that were proven in general above are borne out of this controlled approximation as illustrated in Fig.(\ref{fig2}). First, panels (c), (e), and (f) clearly identify gapless excitations, so our insulating ground state is not protected, and the system is therefore not an insulator. Panels (b), (c), (e), and (f) indicate that the decoupling remains as evidenced by its symptom of a large degeneracy, so the system is not a 3D superfluid. Finally, panel (d) identifies a linear DOS at low energy, which stabilizes the system at finite temperature and proves that this result is not simply a zero-temperature idealization.

Similarly, this approximate solution also produces zero superfluid density as expected from the general proof above. To that end, we use the following thought experiment to identify the current-current response in the low-frequency, long-wavelength limit. Two conducting leads are attached to a sample in an arbitrary orientation as in Fig.(\ref{expt}). These leads are centered at $\mathbf{x}$ and $\mathbf{x'}$ such that the number of lattice spacings $a$ between them is large ($r=|\mathbf{x}-\mathbf{x'}| \gg a$). The size of the leads themselves span many lattice spacings as well, but far fewer than the distance between the leads. Long-range coherent transport is achieved when the current-current response from one lead to the other converges to a finite value for arbitrarily large $r$ and arbitrarily small frequency $\omega$. The conductivity can be evaluated via

\begin{align}
\label{coh}
\sigma(\mathbf{x}-\mathbf{x'},\omega) &= \frac{1}{\omega} \text{Im} \chi^{ll}(\mathbf{x}-\mathbf{x'},\omega),
\end{align}
where $\chi^{ll}(\mathbf{x}-\mathbf{x'},\omega)$ is the longitudinal component of the current-current response function at frequency $\omega$ between points $\mathbf{x}$ and $\mathbf{x'}$. In the supplementary material, Eq.(\ref{coh}) is shown to reproduce the correct superfluid contribution in the unfrustrated regime $\tau'>\tau''$ as expected. 

In the frustrated regime $\tau''>\tau'$ our choice of overall phases for the slabs given by Eq.(\ref{pipiphz}) once again provides an upper bound on the superfluid density. We find that at zero temperature the superfluid contribution $\sigma(r,\omega = 0) = 0$. 

This result can be readily understood from the disconnected nature of the slabs at low energy. Consistent with our general proof above, the hopping parameter $\tau'$ that couples neighbor coherent slabs is suppressed at low energy within this controlled approximation. As a result, any transport that crosses through successive coherent slabs is also suppressed. As demonstrated in Fig.(\ref{expt}), no coherent slabs simultaneously intersect, for example, the source at $\mathbf{x}$ and the sink at $\mathbf{x'}$. Therefore, the DC conductivity vanishes between $\mathbf{x}$ and $\mathbf{x'}$, which includes both the superfluid and normal DC response at zero temperature. 

At finite temperature the normal contribution will develop a finite value given the fact that the gapless excitations at finite energy have weak but finite coherence. More generally, we have proven above and in the supplementary material that this coherence grows faster than exponential with increasing energy. Therefore the zero temperature insulating behavior fails at any finite temperature and a metal obtains.

\subsection*{Other Considerations and Implications}

Although the dressed particle is localized in a single slab due to geometric frustration, this physics is qualitatively unrelated to Anderson localization. In the latter, randomness causes localization and creates an energy scale that must be overcome to reach a channel for transport. If we remove the randomness then we remove this scale and the localized states. In our case, the localization is caused by perfect geometric frustration, which is unrelated to randomness and should be regarded as a high energy constraint beyond the scale of the Hamiltonian. Any state other than this particular many-body ground state is unprotected by perfect frustration and is not localized. The perfect geometric frustration is protected by symmetry, so even when we choose overall phases for each slab that appear coherent as in our above demonstration, each particle in the ground state remains confined to the slab and the superfluid density vanishes. 

On the other hand, full coherence can in principle be recovered via disruption of the perfect interference, for example, by introducing anisotropy in $\tau'$. In that case, the ground state would host a superfluid with $p$-wave symmetry. Accepting the claim\cite{yildirimWeiku} that this model is representative for the extreme low-doping ($<5\%$) non-superconducting regime of the cuprates, this implies that applying enough uniaxial pressure along the (110) crystallographic direction will generate a second superconducting dome in the phase diagram. This dome is separated from the original dome of $d$-wave superfluidity by a quantum critical point due to the difference in their local symmetry.

\section*{Conclusions}

By introducing perfect geometrical frustration among many adjacent nearly free flowing subsystems, we have discovered a long-sought universal stable Bose metal phase intervening between superfluid and insulator. Utilizing a gapless failed insulator in the presence of temperature or disorder immediately leads to dissapative transport. We demonstrate this concept via a 2-band Bose-Hubbard model in an extended checkerboard lattice with frustrated coupling between vertical slabs and a particle density that scales with the system volume. We find that each dressed particle in the many-body ground state is confined to a single slab such that the low-frequency conductivity and the superfluid response both vanish in a general direction at zero temperature. At finite temperature, the system is stable due to the suppressed density of states at low energy. The universal mechanism of our Bose metal leads to a stable phase of quantum matter that is robust under realistic conditions and should therefore be ubiquitous in nature. Engineering this Bose metal in the lab (using ultra-cold atomic gasses for example) and explaining otherwise mysterious metallic behavior plaguing many prototypical strongly correlated materials are just a few of the exciting possibilities that the discovery of this new paradigm entails.

We thank Anthony Leggett for valuable comments concerning our idea. We also thank Jianda Wu and Zi-Jian Lang for helpful discussions as well as Alexei Tsvelik and Jan Zaanen for pointing out relevant studies. This work is supported by National Natural Science Foundation of China Grants 11674220 and 12042507.

\typeout{get arXiv to do 4 passes: Label(s) may have changed. Rerun}

\bibliography{MainTex}

\clearpage
\pagebreak
\widetext
\begin{center}

\textbf{\large Supplemental Materials: Geometric frustration produces long-sought Bose metal phase of quantum matter}
\end{center}
\begin{center}
\text{Anthony Hegg, Jinning Hou, and Wei Ku}
\end{center}
\setcounter{equation}{0}
\setcounter{figure}{0}
\setcounter{table}{0}
\setcounter{page}{1}
\makeatletter
\renewcommand{\theequation}{S\arabic{equation}}
\renewcommand{\thefigure}{S\arabic{figure}}
\renewcommand{\bibnumfmt}[1]{[S#1]}
\renewcommand{\citenumfont}[1]{S#1}

\section{Lack of Flow in General Directions at Low Energy}
\label{TheProof}

\subsection{Checkerboard Lattice Hamiltonian}

\begin{figure}[ht!]
\centering
\includegraphics[scale=0.2]{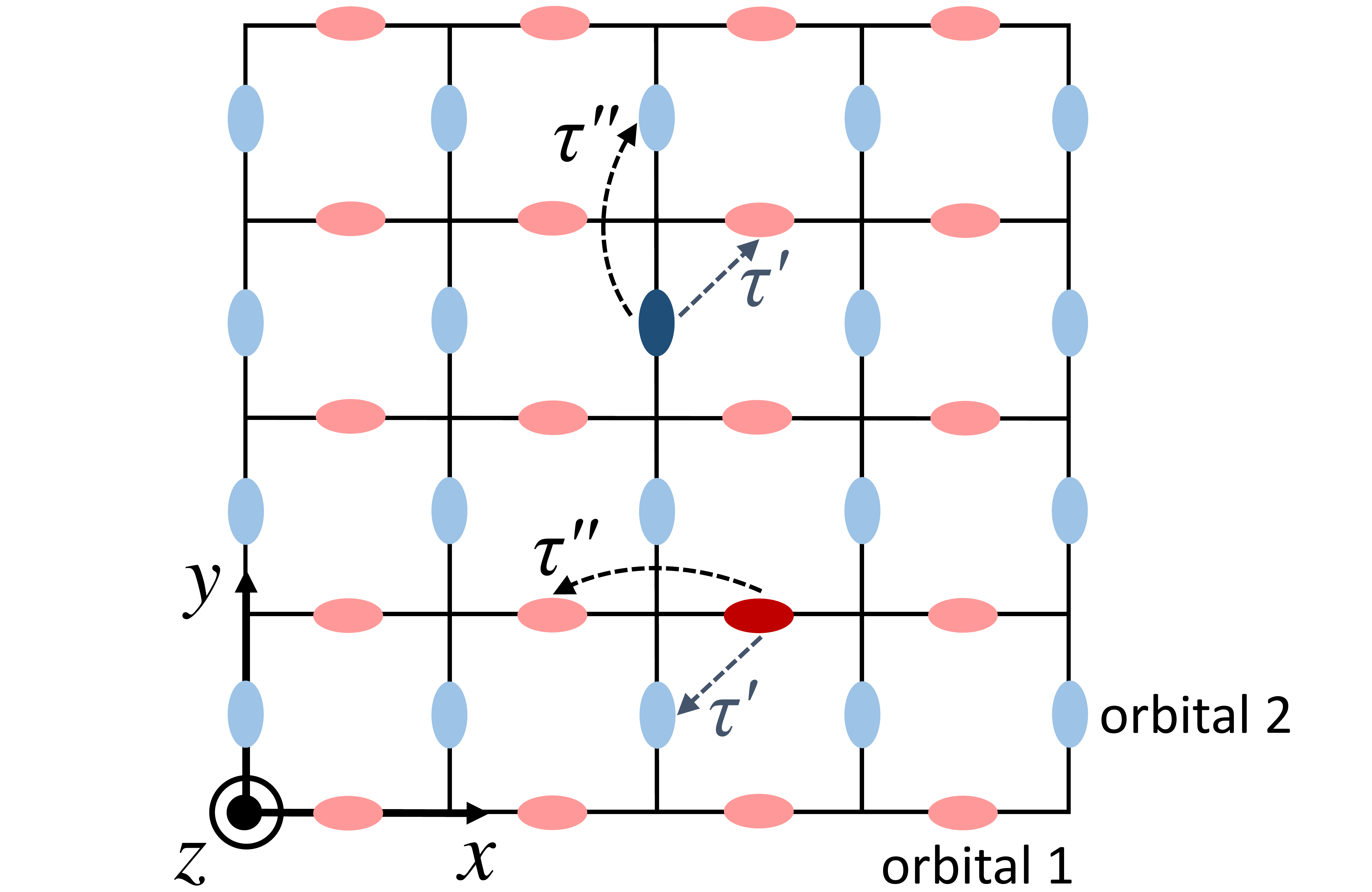}
\caption{The checkerboard lattice with nearest neighbor hopping $\tau'>0$ between sublattices and restricted next-nearest neighbor hopping $\tau''>0$ within the same sublattice. The $z$-axis consists of simple stacking of identical layers with nearest neighbor hopping $\tau_{z}<0$ within the same sublattice. We study the regime where $\tau' < \tau''$, and note that in the absence of $\tau'$ the system decouples into independent $xz$- and $yz$-slabs for orbitals 1 (red) and 2 (blue) respectively. We prove that the fully interacting many-body ground state of Eq.(\ref{chkbd}) maintains this effective decoupling.}
\label{suppfig1}
\end{figure}

The extended checkerboard lattice model for interacting bosons can be written

\begin{align}
\label{chkbd}
\mathcal{H} &= \sum_{i}\Big\{ \sum_{j}\tau_{ij} a^{\dag}_{j} a_{i} + U a^{\dag}_{i} a^{\dag}_{i} a_{i} a_{i} \Big\},
\end{align}
where $i$ is a lattice site with $j$ neighboring bonds and $U>0$. The hopping for this lattice, illustrated in Fig.(\ref{suppfig1}), is given by

\begin{align}
\label{hops1}
\tau_{ij} &= \tau', j \in \{ i + (\pm\frac{1}{2},\pm\frac{1}{2},0) \} \cup \{ i + (\pm\frac{1}{2},\mp\frac{1}{2},0) \}  ~\forall~i \\
\label{hops2}
\tau_{ij} &= \tau'', j \in  \{i + (\pm1,0,0) \} \text{ for } i\in \text{orbital 1,} \text{ and } j \in \{i + (0,\pm1,0) \} \text{ for } i\in \text{orbital 2} \\
\label{hopsz}
\tau_{ij} &= \tau_{z}, j \in  \{i + (0,0,\pm1) \} ~\forall~i,
\end{align}
where the unit vectors are defined by the primitive cell and orbitals $1$ (red) and $2$ (blue) denote the sublattices with horizontal and vertical $\tau''$ hopping directions respectively. We assume periodic boundary conditions for $L$ primitive cells along each direction for a total of $V=2L^3$ lattice sites. Throughout this work we focus on the regime where $\tau'' > \tau' > 0 > \tau_z$. The in-plane positive hopping values favor $\pi$ phase changes across each bond, which puts $\tau'$ and $\tau''$ in direct frustrated competition. 

Our goal is to eventually evaluate the superfluid response using the long-wavelength (momentum $q \rightarrow 0$ limit) of the longitudinal current-current response function. To that end, we identify the current operator in the standard way (to preserve the equation of continuity)

\begin{align}
\label{genCurrOp}
    \vec{\mathcal{J}} &= i\sum_{ij} \vec{\tau}_{ij} \Big( a^{\dag}_{j} a_{i} - a^{\dag}_{i} a_{j} \Big) \equiv \vec{\mathcal{J'}} + \vec{\mathcal{J''}} + \vec{\mathcal{J}_{z}},
\end{align}
where $\vec{\tau}_{ij}$ is a vector of magnitude $\tau_{ij}$ in the direction of the hopping from site $i$ to site $j$. 

Below we will show that the local point group symmetry of $\mathcal{H}$ in Eq.(\ref{chkbd}) can be used to identify a special eigenoperator set that is \emph{always} effectively disconnected in the one-body channel. When $\tau'' > \tau' > 0$ and the interactions $U>0$ are not too strong, these special states are energetically favored and dominate the many-body ground state. As a result of this effectively disconnected ground state structure, we show that there is no superfluid response between slabs (each defined by the $\tau''$ and $\tau_{z}$ lattice) and therefore no superfluid response in the bulk system.

\subsection{Local Point Group Parity}

\begin{figure}[ht!]
\centering
\includegraphics[scale=0.4]{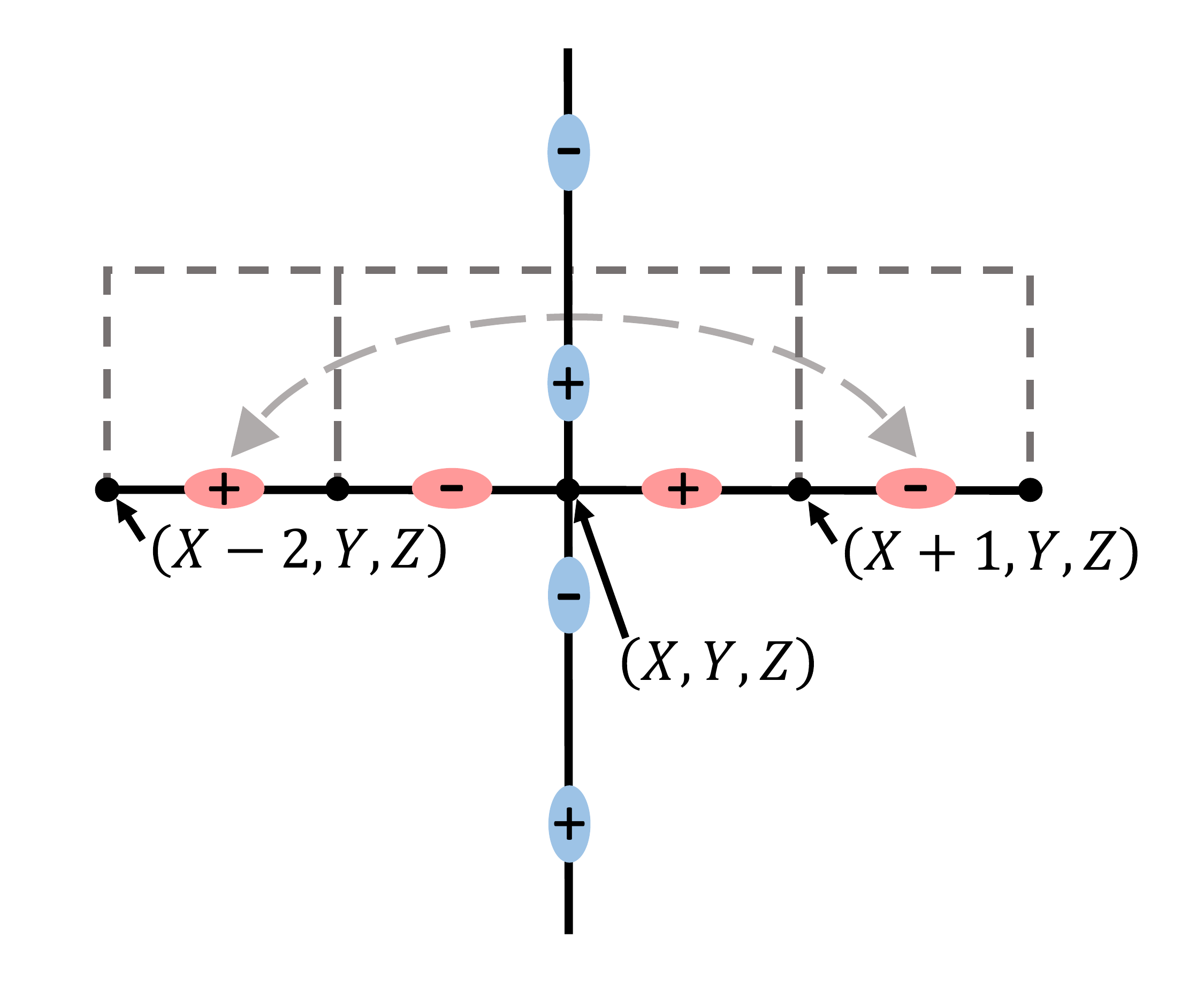}
\caption{Alternating in-plane phase structure located at one $xz$-slab (red) and one $yz$-slab (blue) centered at the point $\mathbf{R}$ where they cross within primitive cell outlined in grey. The signs at each orbital are the phases relative to their neighbors within the same slab. This illustration clearly shows that the parity of such a configuration is odd(even) in the direction parallel(perpendicular) to the slab. For example, the horizontal orbital in unit cell $X+1$ is located at $X+1+\frac{1}{2}$ and, through parity $P^{x}_{XY}$, this orbital maps to the horizontal orbital in unit cell $X-2$ and is located at $X-1-\frac{1}{2}$, which is of opposite phase.}
\label{suppfig2}
\end{figure}

Let's first establish an important local point group symmetry of the system at each lattice site $(X,Y,Z)$, where

\begin{align}
\label{Cells1}
\mathbf{r} &= \mathbf{R} + \mathbf{r}_{\alpha} = (x,y,z) \in \mathbb{R}^3 \\
\mathbf{R} &= (X,Y,Z) \in \mathbb{Z}^3 \\
\label{Cells2}
\mathbf{r}_{1} &= (\frac{1}{2},0,0) \\ 
\label{Cells3}
\mathbf{r}_{2} &= (0,\frac{1}{2},0)
\end{align}
as illustrated in Fig.(\ref{suppfig2}). We define the local point group parity transformation in the horizontal ($P^{x}_{XY}$) and vertical ($P^{y}_{XY}$) directions at the intersection $(X,Y)$ between two slabs via

\begin{align}
    \label{PTransf1}
    P^{x \dag}_{XY} a^{\dag}_{1(X+X',Y,Z)} P^{x}_{XY} &= a^{\dag}_{1(X-X'-1,Y,Z)} \\
    \label{PTransf2}
    P^{x \dag}_{XY} a^{\dag}_{2(X,Y+Y',Z)} P^{x}_{XY} &= a^{\dag}_{2(X,Y+Y',Z)} \\
    \label{PTransf3}
    P^{y \dag}_{XY} a^{\dag}_{1(X+X',Y,Z)} P^{y}_{XY} &= a^{\dag}_{1(X+X',Y,Z)} \\
    \label{PTransf4}
    P^{y \dag}_{XY} a^{\dag}_{2(X,Y+Y',Z)} P^{y}_{XY} &= a^{\dag}_{2(X,Y-Y'-1,Z)}.
\end{align}
For example, Eq.(\ref{PTransf1}) is apparent considering that the horizontal orbital in unit cell $X+X'$ is located at $X+X'+\frac{1}{2}$ and, through parity $P^{x}_{XY}$, it maps to the horizontal orbital in unit cell $X-X'-1$ and is located at $X-X'-\frac{1}{2}$.

Applying Eqs.(\ref{PTransf1}-\ref{PTransf4}) sequentially to an arbitrary local operator $a^{\dag}$ we find

\begin{align}
\label{pxycomm}
    [P^{x \dag}_{XY},P^{y}_{X'Y'}] &= 0 \\
\label{pxxcomm}
    [P^{x \dag}_{XY},P^{x}_{X'Y'}] &\neq 0 \text{ unless } X'=X \text{ and } Y'=Y.
\end{align}
Crucially, the Hamiltonian in Eq.(\ref{chkbd}) commutes with these parity operators

\begin{align}
\label{pxHcomm}
    P^{x \dag}_{XY} \mathcal{H} P^{x}_{XY} &= \mathcal{H} \\
\label{pyHcomm}
    P^{y \dag}_{XY} \mathcal{H} P^{y}_{XY} &= \mathcal{H},
\end{align}
so the eigenoperators of $P^{x}_{XY}$ and $P^{y}_{XY}$ can diagonalize the Hamiltonian as well. However, since parity operators at different $(X,Y)$ do not commute in general, we study $P^{x}_{XY}$ and $P^{y}_{XY}$ for a fixed $X$ and $Y$ below.

\subsection{Example Eigenoperator of Parity}

For clarity, we demonstrate the parity operations defined above for fixed $X$ and $Y$ via the one-body operators

\begin{align}
\label{adag1}
    a^{\dag}_{1,\pi,Y,0} &\equiv \frac{1}{\sqrt{L^2}} \sum_{XZ} e^{i\pi (\mathbf{R}+\mathbf{r}_{1}) \cdot \hat{x}} a^{\dag}_{1\mathbf{R}} = \frac{1}{\sqrt{L^2}} \sum_{XZ} e^{i\pi (X + \frac{1}{2})} a^{\dag}_{1\mathbf{R}}\\
    \label{adag2}
    a^{\dag}_{2,X,\pi,0} &\equiv \frac{1}{\sqrt{L^2}} \sum_{YZ} e^{i\pi (\mathbf{R}+\mathbf{r}_{2}) \cdot \hat{y}} a^{\dag}_{2\mathbf{R}} = \frac{1}{\sqrt{L^2}} \sum_{YZ} e^{i\pi (Y+\frac{1}{2})} a^{\dag}_{2\mathbf{R}},
\end{align}
corresponding to those illustrated in Fig.(\ref{suppfig2}). Applying the parity transformations in Eqs.(\ref{PTransf1}-\ref{PTransf4}) to the operators defined in Eqs.(\ref{adag1},\ref{adag2}) we find

\begin{align}
\label{par1}
    P^{x \dag}_{XY} a^{\dag}_{1,\pi,Y,0} P^{x}_{XY} &= (-1) a^{\dag}_{1,\pi,Y,0} \\
    \label{par2}
    P^{x \dag}_{XY} a^{\dag}_{2,X,\pi,0} P^{x}_{XY} &= (+1) a^{\dag}_{2,X,\pi,0} \\
    \label{par3}
    P^{y \dag}_{XY} a^{\dag}_{1,\pi,Y,0} P^{y}_{XY} &= (+1) a^{\dag}_{1,\pi,Y,0} \\
    \label{par4}
    P^{y \dag}_{XY} a^{\dag}_{2,X,\pi,0} P^{y}_{XY} &= (-1) a^{\dag}_{2,X,\pi,0}.
\end{align}

For example, we obtain Eq.(\ref{par1}) via
\begin{align}
    P^{x \dag}_{XY} a^{\dag}_{1,\pi,Y,0} P^{x}_{XY} &=  \frac{1}{\sqrt{L^2}} \sum_{X'Z'} e^{i\pi (X + X' + \frac{1}{2})} P^{x \dag}_{XY} a^{\dag}_{1(X+X',Y,Z')} P^{x}_{XY} \\
    &= \frac{1}{\sqrt{L^2}} \sum_{X'Z'} e^{i\pi (X + X' + \frac{1}{2})} a^{\dag}_{1(X-X'-1,Y,Z')} \\
    &= \frac{1}{\sqrt{L^2}} \sum_{X''Z'} e^{i\pi (X - X'' - 1 + \frac{1}{2})} a^{\dag}_{1(X+X'',Y,Z')} \\
    &= (-1) \frac{1}{\sqrt{L^2}} \sum_{X''Z'} e^{i\pi (X + X'' + \frac{1}{2})} a^{\dag}_{1(X+X'',Y,Z')} \\
    &= (-1) a^{\dag}_{1,\pi,Y,0},
\end{align}
where $X''=-X'-1$ and $e^{i\pi X'}=e^{-i\pi X'}$ since $X'$ is an integer. As we will show below, the states in this example will play an important role in identifying the symmetry properties of the fully interacting many-body ground state and the subsequent superfluid response of the system.

\subsection{Decomposition of One-body Operator Space into Four Sectors}

Since the parity operators defined at a fixed point $(X,Y)$ commute with each other as well as the Hamiltonian via Eqs.(\ref{pxycomm}-\ref{pyHcomm}), we can define and split the complete one-body operator basis into four distinct sectors based on their $(P^{x}_{XY},P^{y}_{XY})$ eigenvalue pairs. We label any state from each sector via

\begin{align}
    a^{\dag}_{(o,o)} &= \{a^{\dag}_{\lambda} : (P^{x \dag}_{XY} a^{\dag}_{\lambda} P^{x}_{XY},P^{y \dag}_{XY} a^{\dag}_{\lambda} P^{y}_{XY}) = (-a^{\dag}_{\lambda},-a^{\dag}_{\lambda}) \} \\
    a^{\dag}_{(e,o)} &= \{a^{\dag}_{\lambda} : (P^{x \dag}_{XY} a^{\dag}_{\lambda} P^{x}_{XY},P^{y \dag}_{XY} a^{\dag}_{\lambda} P^{y}_{XY}) = (+a^{\dag}_{\lambda},-a^{\dag}_{\lambda}) \} \\
    a^{\dag}_{(o,e)} &= \{a^{\dag}_{\lambda} : (P^{x \dag}_{XY} a^{\dag}_{\lambda} P^{x}_{XY},P^{y \dag}_{XY} a^{\dag}_{\lambda} P^{y}_{XY}) = (-a^{\dag}_{\lambda},+a^{\dag}_{\lambda}) \} \\
    a^{\dag}_{(e,e)} &= \{a^{\dag}_{\lambda} : (P^{x \dag}_{XY} a^{\dag}_{\lambda} P^{x}_{XY},P^{y \dag}_{XY} a^{\dag}_{\lambda} P^{y}_{XY}) = (+a^{\dag}_{\lambda},+a^{\dag}_{\lambda}) \},
\end{align}
where $a^{\dag}_{\lambda}$ is a simultaneous eigenoperator of $P^{x}_{XY}$, $P^{y}_{XY}$, and $\mathcal{H}$.

For example, from Eqs.(\ref{par1}-\ref{par4}) we identify the following inclusions

\begin{align}
\label{1py}
a^{\dag}_{1,\pi,Y,0} \in a^{\dag}_{(o,e)} \\
\label{2xp}
a^{\dag}_{2,X,\pi,0} \in a^{\dag}_{(e,o)}.
\end{align}
On the other hand, for operators with uniform parity along each slab direction, we find

\begin{align}
\label{10y}
a^{\dag}_{1,0,Y,0} \in a^{\dag}_{(e,e)} \\
\label{1x0}
a^{\dag}_{2,X,0,0} \in a^{\dag}_{(e,e)}.
\end{align}
Note that the examples presented here in Eqs.(\ref{1py}-\ref{1x0}) are special in that they contain a very high degree of symmetry. Specifically, they are the only states in their respective slabs that belong to the same parity sector for \emph{all} $(X,Y)$ points in their respective slabs, where each horizontal(vertical) slab is defined by its vertical(horizontal) location $Y$($X$). Interestingly, these states are simultaneous eigenstates of all $(P^{x}_{XY},P^{y}_{XY})$ in their respective slabs despite the fact that these parity operators do not commute in general (cf. \ref{pxxcomm}).

\subsection{Fully Dressed One-Body Operators and their Parity Properties}

Since our goal is to investigate the superfluid response, Eq.(\ref{genCurrOp}) suggests that it will be most convenient to represent the Hamiltonian Eq.(\ref{chkbd}) in its diagonal form using the `fully dressed' one-body operator $\tilde{a}^{\dag}$ via

\begin{align}
\label{HDiag}
 \mathcal{H}[a^\dag,a] = \mathcal{H}_{D}[\tilde{a}^{\dag},\tilde{a}] &= \sum_{I} E_{I} \tilde{a}^{\dag}_{I} \tilde{a}_{I} + \sum_{IJ} E_{IJ} \tilde{a}^{\dag}_{I} \tilde{a}^{\dag}_{J} \tilde{a}_{J} \tilde{a}_{I} + \cdots \\
 \label{UTransf}
 \tilde{a}^{\dag}_{I} &= \mathcal{U}^{\dag} a^{\dag}_{i} \mathcal{U} ~~~(I \leftrightarrow i),
\end{align}
where $\tilde{a}^{\dag}_{I}$ creates a particle in a fully-dressed (by many-body processes) one-body state with index $I$ and $E_{I}$, $E_{IJ}$, etc. are the one-body, two-body, etc. partitions of the many-body energy eigenstates respectively up to $N$-body. Note the one-to-one mapping between $I$ and $i$ indices. Typically, $I$ would automatically be a momentum $k$ due to translational symmetry. However, in cases of localization it can also be a real space site index $(X,Y)$. 

Since $\mathcal{U}$ is responsible for diagonalizing the Hamiltonian, which commutes with the parity operators, transformation via $\mathcal{U}^{\dag}$ and $\mathcal{U}$ must preserve the symmetry. In particular, since we can choose the set of bare operators $\{ a^{\dag} \}$ to be simultaneous eigenoperators of $P^{x}_{XY}$, $P^{y}_{XY}$, and $\mathcal{H}$, the unitary transformation in Eq.(\ref{UTransf}) preserves this relation and we find

\begin{align}
    \label{UTT1}
    \tilde{a}^{\dag}_{(o,o)} &= \mathcal{U}^{\dag} a^{\dag}_{(o,o)} \mathcal{U} \\
    \label{UTT2}
    \tilde{a}^{\dag}_{(e,o)} &= \mathcal{U}^{\dag} a^{\dag}_{(e,o)} \mathcal{U} \\
    \label{UTT3}
    \tilde{a}^{\dag}_{(o,e)} &= \mathcal{U}^{\dag} a^{\dag}_{(o,e)} \mathcal{U} \\
    \label{UTT4}
    \tilde{a}^{\dag}_{(e,e)} &= \mathcal{U}^{\dag} a^{\dag}_{(e,e)} \mathcal{U}.
\end{align}
In other words, Eqs.(\ref{UTT1}-\ref{UTT4}) indicate that the set of fully-dressed one-body operators that generate the many-body energy eigensolution of the Hamiltonian in Eq.(\ref{HDiag}) inherit the parity structure of the bare one-body operators.

\subsection{Special Fully-Dressed One-Body State with High Symmetry}

We now look at the operators defined in Eqs.(\ref{adag1},\ref{adag2}), illustrated in Fig.(\ref{suppfig2}) and further transformed

\begin{align}
\label{piXTransf}
    \tilde{a}^{\dag}_{I,\pi} &\equiv \mathcal{U}^{\dag} a^{\dag}_{2,X,\pi,0}  \mathcal{U} ~~~(I \leftrightarrow X) \\
\label{piYTransf}
    \tilde{a}^{\dag}_{\pi,J} &\equiv \mathcal{U}^{\dag} a^{\dag}_{1,\pi,Y,0}  \mathcal{U} ~~~(J \leftrightarrow Y)
\end{align}
via Eqs.(\ref{UTT2},\ref{UTT3}) respectively, which we denote as  $\tilde{a}^{\dag}_{I,\pi}$ and $\tilde{a}^{\dag}_{\pi,J}$ with indices $I$ and $J$ (cf. \ref{UTransf}) for vertical and horizontal odd parity sectors respectively. In terms of the parity symmetry operators $P^{x}_{XY}$ and $P^{y}_{XY}$ defined above, $\tilde{a}^{\dag}_{\pi,J}$ share the unique property that each has odd parity along the slab direction for \emph{every} $(X,Y)$ point in its corresponding slab, even though the parity operators at different $(X,Y)$ points do not commute. In contrast, all of the other states in this system have momentum $k$ as their good quantum number, as typically expected of a lattice with translational symmetry. 

The especially high degree of symmetry here leads to an interesting $N$-body state

\begin{align}
    \label{PiState}
    \ket{\Pi} &\equiv \prod_{IJ} \Big( \tilde{a}^{\dag}_{I,\pi} \Big)^{N_{I}} \Big( \tilde{a}^{\dag}_{\pi,J} \Big)^{N_{J}} \ket{0} \\
    N &= \sum_{I} N_{I} + \sum_{J} N_{J},
\end{align}
which inherits the especially high symmetry in the $\tilde{a}_{\pi,J}$'s. All particles in orbital $1$ belonging to the fully-dressed $(\pi,J)$ state cannot propagate to orbital $2$

\begin{align}
    \bra{\Pi} a^{\dag}_{2,X,Y',Z} \tilde{a}_{\pi,J(Y)} \ket{\Pi} &= \bra{\Pi} P^{x}_{XY} \Big( P^{x \dag}_{XY}  a^{\dag}_{2,X,Y',Z} P^{x}_{XY} \Big) \Big(P^{x \dag}_{XY} \tilde{a}_{\pi,J(Y)} P^{x}_{XY} \Big) P^{x \dag}_{XY} \ket{\Pi} \\
    &= \bra{\Pi} (+1) a^{\dag}_{2,X,Y',Z} (-1) \tilde{a}_{\pi,J(Y)} \ket{\Pi} \\
    &= (-1) \bra{\Pi} a^{\dag}_{2,X,Y',Z} \tilde{a}_{\pi,J(Y)} \ket{\Pi} \\
    \label{ZeroOv}
    \Rightarrow \bra{\Pi} a^{\dag}_{2,X,Y',Z} \tilde{a}_{\pi,J(Y)} \ket{\Pi} &= 0,
\end{align}
where $a_{\pi,J(Y)}$ annihilates a dressed particle transformed from $a_{1,\pi,Y,0}$ and $a^{\dag}_{2,X,Y',Z}$ creates a particle in any orbital $(2,X,Y',Z)$, which is located in a vertical slab. In other words, all one-body particles in this $N$-body state are confined to move only in orbital $1$ or only in orbital $2$

\begin{align}
\label{MobZero}
    \bra{\Pi} a^{\dag}_{2,X,Y,Z} a_{1,X',Y',Z'} \ket{\Pi} &= 0.
\end{align}

\subsection{Geometrical Frustration and One-Body Localization}
\label{GeoFrus}

The above limited mobility of the particles in these special $\ket{\Pi}$ states, when combined with the geometrical frustration of the system, can lead to an interesting one-body localization. For example, in the Hamiltonian of Eq.(\ref{chkbd}), a particle in orbital $1$ from a particular $Y$ slab can only propagate out of the slab via $\mathcal{H}_{\tau'}$ to the neighboring orbital $2$ located in a vertical slab. When symmetry prevents particles in $\ket{\Pi}$ from propagating to half of the orbitals, $\mathcal{H}_{\tau'}$ is effectively disabled ($\bra{\Pi}\mathcal{H}_{\tau'}\ket{\Pi}=0$ via Eq.(\ref{MobZero})), and therefore each particle is effectively localized to a single slab (one-body localization). As a result, for these states the `slab indices' $X$ and $Y$ can be chosen to label $\tilde{a}^{\dag}_{I,\pi}$ and $\tilde{a}^{\dag}_{\pi,J}$ as $\tilde{a}^{\dag}_{X,\pi}$ and $\tilde{a}^{\dag}_{\pi,Y}$ respectively. Such a one-body localization is a generic feature of geometrically frustrated many-body systems. 

\begin{figure}[ht!]
\centering
\includegraphics[scale=0.2]{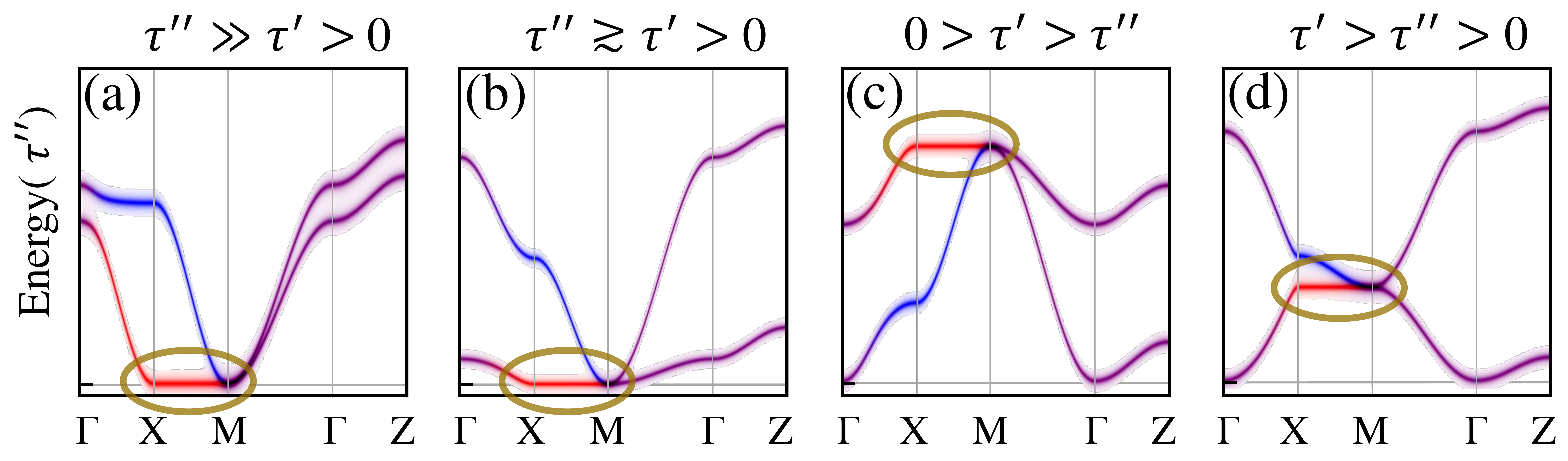}
\caption{The dispersion of $\mathcal{H}$ in Eq.(\ref{chkbd}) for several qualitatively different hopping regimes. Note that the line degeneracy remains precisely intact regardless of these parameters. This property is a consequence of the lattice symmetry, which is why it will remain intact regardless of the presence of local density-density interactions $U$.}
\label{dispsupp}
\end{figure}

Such symmetry protection of localization is easily observed in Fig.(\ref{dispsupp}).
In the absence of $U$, the one-body propagator displays a perfectly flat dispersion (circled in each panel) along a straight line in momentum space \emph{independent} of parameters $\tau'$ and $\tau''$. This robust, parameter-independent line degeneracy reflects the underlying symmetry protection of such a one-body localization.

\begin{figure}[ht!]
\centering
\includegraphics[scale=0.5]{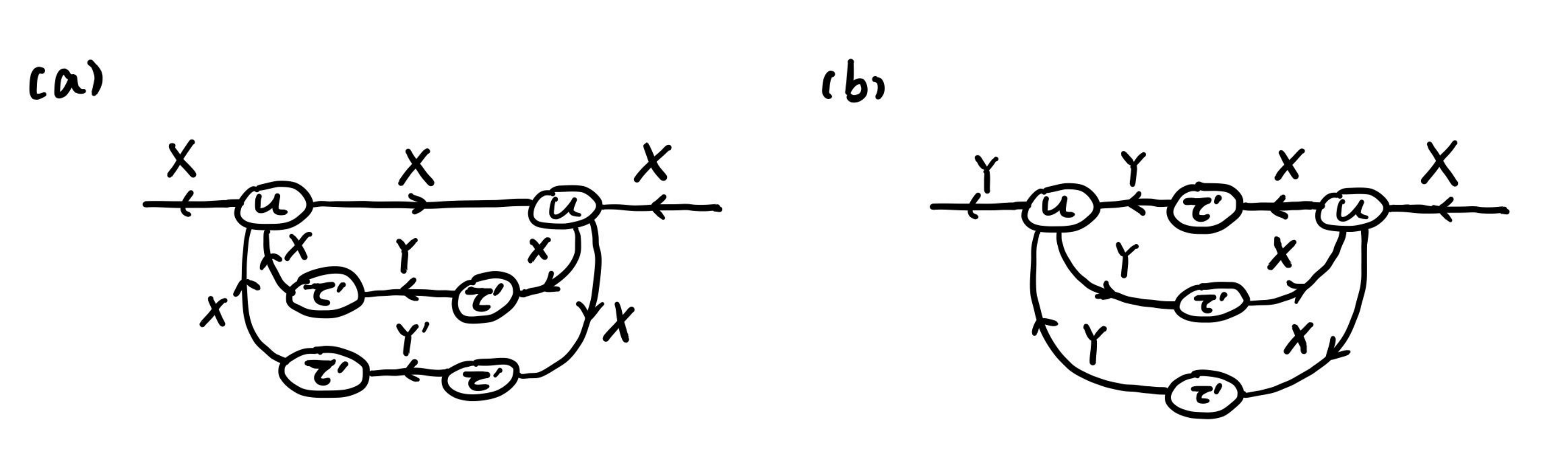}
\caption{Examples of virtual processes in the interacting system for $\tilde{a}^{\dag}$. (a) A virtual process in which $\tilde{a}^{\dag}$ remains in the same initial slab $X$ while in the presence of many-body quantum fluctuations involving the neighboring slab $Y$. (b) A virtual process in which $\tilde{a}^{\dag}$, initially in slab $X$, propagates to slab $Y$ while in the presence of many-body quantum fluctuations. While this process appears active for all one-body states, Eq.(\ref{ZeroOv}) proves that such diagrams must sum to zero for the special states $\tilde{a}^{\dag}_{X,\pi}$.}
\label{suppdiagrams}
\end{figure}

Since the symmetry protection discussed above has a (non-local) geometric origin, it is no surprise that the symmetry-protected result of Eq.(\ref{ZeroOv}) remains robust in the presence of local density-density interactions $U$. This result explains why, even in the controlled approximation outlined below, the flat dispersion remains unaffected, since this approximation of our many-body system properly respects the underlying lattice symmetry. Of course, one would expect in general that a many-body dressing will generate a cloud of quantum fluctuations around each slab $X$ as illustrated in Fig.(\ref{suppdiagrams})(a). On the other hand, the effect of symmetry protection might not be apparent by listing the possible many-body processes alone (e.g. Fig(\ref{suppdiagrams})(b)), since such protection often manifests itself through perfect cancellation of contributions.

\subsection{Lack of Superfluidity in the Frustrated Regime}

It turns out that $\ket{\Pi}$, defined above in Eq.(\ref{PiState}), is the many-body ground state of our Hamiltonian Eq.(\ref{chkbd}) in the frustrated regime $\tau''>\tau'>0$. This is apparent since Fig.(\ref{dispsupp}) panels (a) and (b) clearly illustrated that the bare $\pi$-states are the ground state for \emph{any} positive $\tau'<\tau''$. If $U$ is not too big, the local interactions will merely dress $\ket{\Pi}$ as the ground state as opposed to promoting a qualitatively distinct ground state structure (e.g. Mott insulating ground state at very large $U$). 

\begin{figure}[ht]
\centering
\includegraphics[scale=0.5]{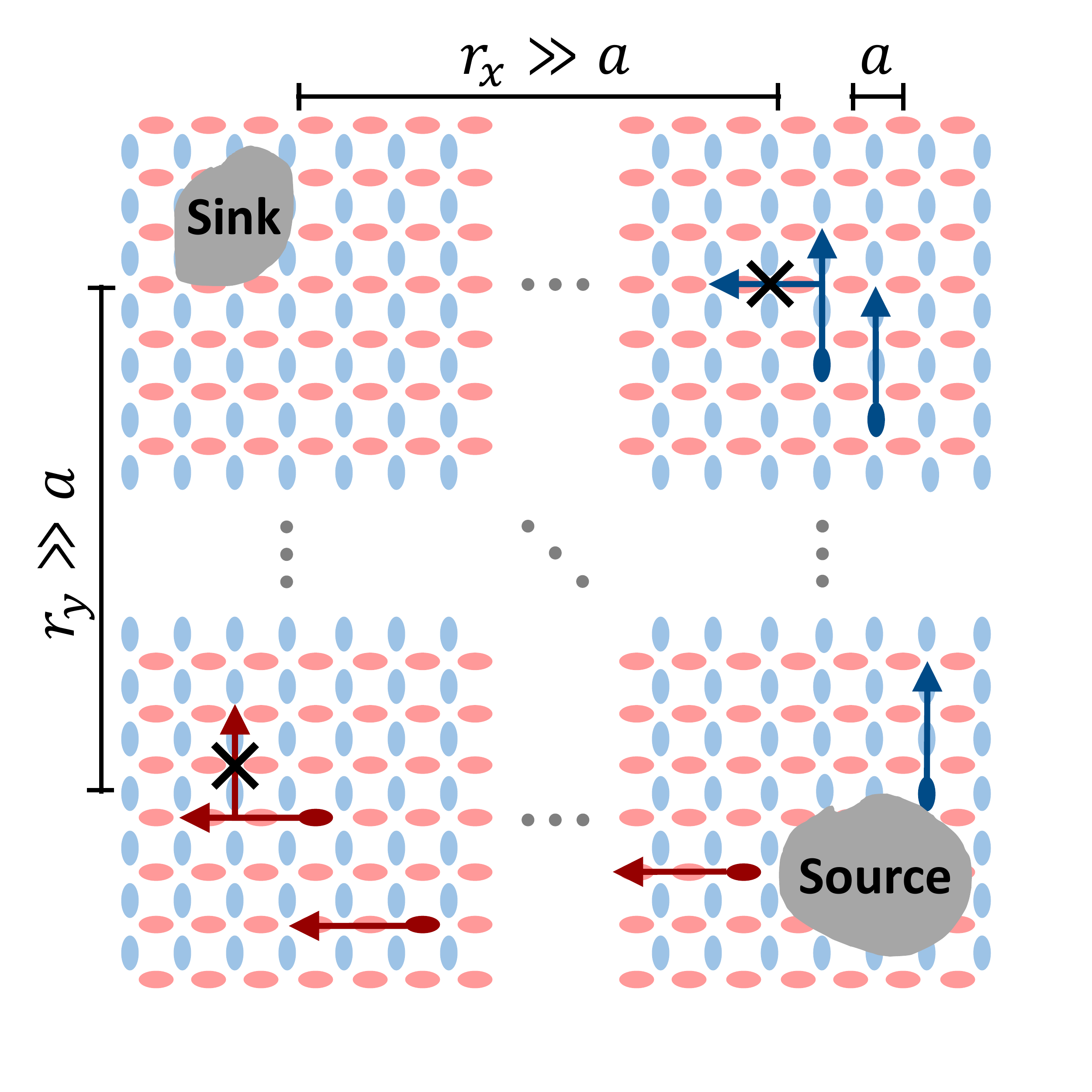}
\caption{Hypothetical experimental probe of the current with two gray contacts separated by a distance $r=\sqrt{r_{x}^2+r_{y}^2}$ much larger than the lattice spacing $a$. Particles emerge from the source and travel in straight lines along a given coherent slab. The probability for the current to flow between sublattices vanishes in the low energy limit, so no path remains to reach the sink if either $r_{x}$ or $r_{y}$ are large enough to place the leads on distinct coherent slabs.}
\label{suppexpt}
\end{figure}

Since the previous subsection demonstrated the one-body localization of $\ket{\Pi}$ and the corresponding lack of propagation between slabs, there can be no superflow. We illustrate this in Fig.(\ref{suppexpt}) with the following thought experiment. Two conducting leads are attached to a sample in an arbitrary orientation . These leads are centered at $\mathbf{R}$ and $\mathbf{R'}$ primitive cells such that the distance $r$ between them is large
\begin{align}
    |\mathbf{r}|=r=\sqrt{r_{x}^2+ r_{y}^2} = |\mathbf{R}-\mathbf{R'}| \gg a,
\end{align}
where we have used the notation in Eqs.(\ref{Cells1}-\ref{Cells3}). The size of the leads themselves spans many lattice spacings $a$ as well, but this distance is much less than the distance between the leads. Long-range coherent transport is achieved when the current-current response from one lead to the other converges to a finite value for arbitrarily large $r$ and arbitrarily small frequency $\omega$. The optical conductivity

\begin{align}
\label{suppcoh}
\sigma (\mathbf{r}) &= \lim_{\omega \rightarrow 0} \frac{1}{\omega} \text{Im} \chi^{ll}(\mathbf{r},\omega)
\end{align}
can be obtained from $\chi^{ll}(\mathbf{r},\omega)$, the longitudinal component of the current-current response function.

As proven above and illustrated in Fig.(\ref{suppexpt}), in the frustrated regime, flow cannot go from horizontal to vertical or vice versa in the low-energy limit. Therefore, there is no low-energy current response at the sink due to applied current at the source, superfluid or otherwise.

\section{Stability of the Many-body Ground State at Low Temperature}

To establish the stability of the system against thermal fluctuations, it is sufficient to confirm that the single particle density of states vanishes linearly at low frequency. We examine this property by first neglecting the $\mathcal{H}_{\tau'}$ as before. Each slab is decoupled, and the low-energy excitations for a slab are well-known (see e.g. \cite{HFMengSupp}) with a linear spectrum and corresponding linear one-body density of states (DOS) in the low-energy limit. 

A system with a DOS that is linear at low energy is stable at low temperature. This can be seen by noting that, for a linear DOS and upon introduction of a small temperature, the chemical potential must be pinned at the one-body band minimum in order to conserve particle number. This property is consistent with the observation of superfluidity in thin films.

We now reintroduce $\mathcal{H}_{\tau'}$, which will in general couple the slabs for high-energy excitations. Such coupling mixes states in a finite energy window and alters the form of the DOS. However, we have already proven that the slabs host one-body localization, and the one-body particles do not propagate. The smoothness of the set of solutions implies that the one-body excitations must become more and more decoupled in the low-energy limit. Therefore the mixing energy window must shrink to zero in the low-energy limit as well. Since the low-energy DOS is already linear, it must remain so since the states that couple in this small window are already linear in this regime, and the details of the coupling do not affect an integrated quantity such as the DOS.

Since the low-energy DOS remains linear after reintroduction of $\mathcal{H}_{\tau'}$, the system is stable at low temperature. In total, we have proven in general that our system of disconnected slabs form a stable phase of bosons at low temperature.

\section{General Properties in the Presence of Disorder}

\subsection{Disorder Induced Coherence Ruled out via Scaling}

Introduction of a single impurity breaks parity locally and establishes a small coupling between extended one-body states in different slabs. We assume that each disordered site located near $(XYZ)$ introduces a term $\epsilon_{XYZ} a^{\dag}_{1XYZ} a_{2XYZ}$ with

\begin{align}
     \epsilon_{XYZ} = \bra{1XYZ} \mathcal{H}_{\tau'} \ket{2XYZ}
\end{align}
being a random variable chosen from a symmetric distribution with zero mean $\braket{\epsilon_{XYZ}}=0$ and nonzero variance $\braket{\epsilon_{XYZ}^2}=\Delta$. For disorder strength weaker than the interaction strength $\Delta \ll U$, the slab (with a finite stiffness) is protected and remains rigid. Therefore, coupling through disorder along the intersection line between two slabs would be $\sum_{XY} V_{XY} a^{\dag}_{\pi,Y,0} a_{X,\pi,0}$ with corresponding matrix element

\begin{align}
\label{VEq}
    V_{XY} &= \frac{1}{L} \sum_{Z} \epsilon_{XYZ},
\end{align}
where the sum is only over impurity sites. Assuming that the number of impurities scale with the system size, the summation in Eq.(\ref{VEq}) includes $\mathcal{O}(L)$ terms. For large $L$, the distribution of the sum scales as $\mathcal{O}(1/\sqrt{L})$ by the central limit theorem, so the overall scaling for this coupling term is bounded by

\begin{align}
\label{VScale}
    V_{XY} &\propto \mathcal{O} \Big( \frac{1}{L\sqrt{L}} \Big).
\end{align}

Since the effect of an impurity and the lack of coherence due to geometrical frustration are both one-body effects, it is sufficient to analyze the energy scale of the induced coherence by analyzing the scaling in a one-body Hilbert space formed by the $2L$ one-body operators $\tilde{a}^{\dag}_{X,\pi}$ and $\tilde{a}^{\dag}_{\pi,Y}$ as defined above. The coupling in Eq.(\ref{VScale}) scales as $\mathcal{O}(1/L\sqrt{L})$, so the leading order correction to the energy scale of the coherence scales as $V_{XY}^2 \sim \mathcal{O}(1/L^3)$. Each slab couples via $V_{XY}$, and there are $L^2$ such intersections. The total relevant energy correction thus scales as $\mathcal{O}(L^2/L^3)=\mathcal{O}(1/L)$, which vanishes in the large system limit. Therefore, this coupling is not thermodynamically meaningful. This analysis does not account for the incoherent nature of disorder, so in that sense it is an upper bound estimate of the coherence generated by impurities. Interested readers are welcome to test this scaling result by setting up an $L \times L$ matrix with zero diagonal elements and dense random off-diagonal elements of order $\mathcal{O}(1/L\sqrt{L})$ and observe that the distribution of the resulting eigenvalues scales to zero in the large $L$ limit. In summary, disorder cannot introduce meaningful 3D coherence of the one-body states in $\ket{\Pi}$, and the slabs remain effectively independent.

\subsection{Stability of Metallicity against Anderson Localization}

In general, the one-body excitations are rigorously coherent in 3D, and the coherence is confined to effective 2D slabs only in the low-energy limit. Representing the low-energy one-body density of states (DOS) as a power law $\omega^{\alpha}$, our stiffened phase system enlarges this power $\alpha=1>1/2>0>-1/2$ relative to 3D, 2D, and 1D normal metals respectively. This indicates that our system should be at least as resilient against Anderson localization as a 3D normal metal if not more so. In addition, the linear dispersion of the low-energy one-body states along each slab implies that they do not have locality in the slab (e.g. their Wannier functions have power-law decay tails). Therefore, their coupling to a local impurity will be strongly suppressed, and we expect in general that localization will require an even larger critical disorder density than what one expects for a normal 3D metal.

\section{Additional Confirmation of both Metallicity and the Lack of Superfluidity}

\subsection{Many-body Reference State Method}

Below, we confirm the metallicity of our Bose metal via a controlled approximation of the many-body solution. First we review the standard reference state method. We transform the operators formally

\begin{align}
\label{denphzform}
a_{i} &= e^{i \hat{\phi}_{i}} \sqrt{\hat{n}_{i}},
\end{align}
where $\hat{n}_{i}=a^{\dag}_{i} a_{i}$ and $\hat{\phi}_{i}$ are assumed to be the Hermitian conjugate operator pair of density and phase operators. The existence of this conjugate operator pair in general has a long and unresolved history, but it is well-defined \cite{RLynchSupp,SperlingVogelSupp} in the case of unbounded bosons

\begin{align}
\hat{n}\ket{m} &= m\ket{m}~~~m \in \{ 0,\pm 1, \pm 2, \ldots \},
\end{align}
where $\ket{m}$ are the complete set of occupation states. The definition of a conjugate operator pair for an operator with discrete eigenvalues $m$ and one with periodic eigenvalues $\phi$ is somewhat subtle\cite{JudgeSupp,KrauseSupp}, but we will use a self-consistent method that avoids these subtleties. 

We can expand these local operators in the fluctuations around a quantum many-body reference state

\begin{align}
    \hat{n}_{i} &= n_{i} + \hat{\eta}_{i} \\
    \hat{\phi}_{i} &= \phi_{i} + \hat{\theta}_{i},
\end{align}
where $n_{i}$ and $\phi_{i}$ are real numbers, and $\hat{\eta}$ and $\hat{\theta}$ are operators. We can then choose a uniform reference density

\begin{align}
    n_{i} &\equiv n_{0},
\end{align}
and if we expand each operator to linear order in the density fluctuations $\eta$ and phase fluctuations $\theta$ we would find

\begin{align}
\label{aexp}
a_{i} &\approx e^{i\phi_{0i}} \sqrt{n_{0}} \left( 1 + \frac{\eta_{i}}{2n_{0}} + i \theta_{i} \right),
\end{align}
where $n_{0}$ and $\phi_{0i}$ are the local reference state values for density and phase at site $i$, and $\eta$ and $\theta$ are the Hermitian density and phase fluctuation operators respectively. 

Confining our study to the regime where the fluctuations are small eliminates the subtleties mentioned previously resulting in the canonical algebra

\begin{align}
\label{denphzalg}
[\eta_{i},\theta_{i'}] &= i\delta_{ii'}.
\end{align}

\subsection{Our Controlled Approximation}

Our method is similar in spirit, however we can instead treat Eq.(\ref{aexp}) as a definition for $\eta$ and $\theta$ i.e.

\begin{align}
\label{atransf}
a_{i} &= e^{i\phi_{0i}} \sqrt{n_{0}} \left( 1 + \frac{\eta_{i}}{2n_{0}} + i \theta_{i} \right),
\end{align}
which clearly satisfies the commutation relation in Eq.(\ref{denphzalg}). This transformation maintains Eq.(\ref{denphzalg}) and the hermiticity of these operators, but we lose the exact physical interpretation of $\eta$ and $\theta$ as density and phase fluctuations respectively. 

Both the expansion and the exact transformation above result in a similar quadratic Hamiltonian, which is the focus of our study, but the exact transformation simplifies a comparison with higher order terms, since there are only a finite number of additional terms in this case. These terms are all in the interaction term $\mathcal{H}_{U}$ and are of the forms $\eta^3$, $\eta^4$, $\eta^2\theta^2$, and $\theta^4$. The low-energy physics is dominated by $\theta$ and the density fluctuations $\eta$ are suppressed ($\eta_k \propto \sqrt{k}$ at the quadratic level for small momentum component $k$). Therefore, the most relevant interactions beyond quadratic are of the form $\theta^4$. For a system with a single lowest energy state (such as in our model for $\tau'>\tau''>0$ or in each disconnected single slab when $\tau''>\tau'>0$), as long as the system is two or higher dimension, these $\theta^4$ correction terms do not alter the low-energy quasiparticle dispersion. Therefore, in the low-temperature limit, the quadratic part of the Hamiltonian captures the essential physics. The smallness of this correction term determines the level of control over errors introduced by this approximation method. The details of this correction term and how it affects the physics as temperature is increased will be published in a future work.

To enforce the internal consistency of these approximations we introduce a Lagrange multiplier $\mu$ via $\mathcal{K} = \mathcal{H} - \mu \sum_{i} (\hat{n}_{i} - n_{0})$ to constrain the average particle number. Expanding $\mathcal{H}=\mathcal{H}_{\text{chkbd}}+ \mathcal{H}_{\text{int}}$ to bilinear order in the fluctuation operators we find

\begin{align}
\label{bilin1}
\mathcal{K} &= \mathcal{H} - \mu \sum_{i} \left( \hat{n}_{i} - n_{0} \right) = \mathcal{K}_{0} + \mathcal{K}_{1} + \mu\sum_{i} n_{0} + \mathcal{T} - \mu \tilde{N} + \mathcal{V} \\
\mathcal{K}_{0} &= \left\{ (2\tau_{z} - 2\tau'') n_{0} +  Un_{0}(n_{0}-1) - \mu n_{0}  \right\} V \\
\mathcal{K}_{1} &= \sum_{i} \left\{ \sum_{j} \tau_{ij} e^{i(\phi_{0i}-\phi_{0j})} \frac{\eta_{i}+\eta_{j}}{2} + \left\{ U(2n_{0}-1)-\mu \right\} \eta_{i} \right\} \\
\mathcal{T} &= \sum_{ij} \tau_{ij} e^{i(\phi_{0i}-\phi_{0j})} \Big\{ \frac{\eta_{i}\eta_{j}}{4n_{0}} + n_{0}\theta_{i}\theta_{j} + \frac{i}{2}\left( \eta_{i}\theta_{j} - \theta_{i}\eta_{j} \right) \Big\}  \\
\tilde{N} &= \sum_{i} \Big\{ \frac{\eta_{i}^2}{4n_{0}} + n_{0}\theta_{i}^2 + \frac{i}{2}\left[ \eta_{i}, \theta_{i} \right] \Big\} \\
\label{bilin6}
\mathcal{V} &= (2n_{0}-1)U\tilde{N} + U \sum_{i} \eta_{i}^2,
\end{align}
where $\mathcal{K}_{0}$ is a constant term independent of fluctuation operators and $V$ is the total number of lattice sites. Note that $\mathcal{K}_{0}$ and $\mathcal{K}_{1}$ are independent of $\tau'$. Enforcing the Lagrange multiplier such that $\braket{\eta}=0$ is satisfied by choosing $\mu$ such that the terms linear in fluctuation operators collectively vanish. If the linear terms do not vanish, then the resulting eigenstates pick up a constant factor representing fluctuation about a reference state that differs from our assumption in e.g. Eq.(\ref{aexp}). The resulting constraint on $\mu$ is

\begin{align}
    0 &= (2\tau_{z} - 2\tau'') +  U(2n_{0}-1) - \mu \\
    \Rightarrow \mu &\rightarrow (2\tau_{z} - 2\tau'') +  U(2n_{0}-1).
\end{align}
Eqs.(\ref{bilin1}-\ref{bilin6}) have a reference state relative phase $\phi_{i}$ at each site that is in principle fixed by e.g. the Hamiltonian. Due to the structure of the checkerboard lattice (c.f. Figs.(\ref{suppfig1},\ref{suppfig2})), when $\tau''>\tau'>0$ there remains an overall phase freedom in the ground state of each slab since they are decoupled.

As an upper bound on the superfluid response, we can artificially choose these phases to have coincident translational invariance with the lattice, e.g. let

\begin{align}
\label{phzx}
    \phi_{1,x,y,z} &= \pi X \\
    \label{phzy}
    \phi_{2,x,y,z} &= \pi Y,
\end{align}
which represents the choice where the two orbitals in a single primitive cell have the same phase, and all other cells are related by a $(\pi,\pi)$ rotation. Certainly, if one were to instead choose a configuration that does not follow the structure of the lattice (e.g. random overall phases for each slab), this would then further disrupt any superfluidity in the system. However, as we show below, even in this upper bound case the superfluid response vanishes in the large system limit.

We are interested in the low-energy behavior of this system, and since we have chosen the phases as in Eqs.(\ref{phzx},\ref{phzy}), the many-body eigenstates are also eigenstates of momentum. So we define

\begin{align}
    \text{\boldmath$\kappa$} &= \mathbf{k} + (\pi,\pi,0) = (\text{\boldmath$\kappa$}_{\text{slab}},\kappa_{\perp})
\end{align}
where we have introduced the following orbital-independent momentum representation

\begin{align}
    \text{\boldmath$\kappa$}_{\text{slab}} = &~(k_{x}+\pi,k_{z})~\text{ for orbital 1} \\
    \nonumber
    &~(k_{y}+\pi,k_{z})~\text{ for orbital 2} \\
    \label{kperp}
    \kappa_{\perp} = &~k_{y}+\pi ~\text{ for orbital 1} \\
    \nonumber
    &~k_{x}+\pi~\text{ for orbital 2}
\end{align}
such that the ground state of each slab is at $\text{\boldmath$\kappa$}_{\text{slab}}=(0,0)$. 

The low-energy eigensolutions are given by

\begin{align}
\label{Hsols}
\mathcal{K}_{\text{eff}} &= \mathcal{K}_{0} + \tilde{\mathcal{K}}_{0} + \sum_{\nu k} E_{\nu k} b^{\dag}_{\nu k} b_{\nu k} \\
b_{\nu k} &= \frac{c_{\nu k}\eta_{\nu k} + d_{\nu k} \eta_{\nu -k} }{2n} +i (c_{\nu k} \theta_{\nu k} - d_{\nu k} \theta_{\nu -k}) \\
E_{\nu k} &= \sqrt{\epsilon_{\nu k} \left( \epsilon_{\nu k} + 4Un_{0} \right)} \\
c_{\nu k} &= \frac{\epsilon_{\nu k} + 2Un_{0} + E_{\nu k}}{2E_{\nu k}} \\
d_{\nu k} &= \frac{\epsilon_{\nu k} + 2Un_{0} - E_{\nu k}}{2E_{\nu k}} \\
\label{Hsolsf}
\tilde{\mathcal{K}}_{0} &= -\frac{1}{2} \sum_{k} \left( \epsilon_{\nu k} + 2Un_{0}- \sqrt{\left(\epsilon_{\nu k} + 2Un_{0}\right)^2 - \frac{U^2 n_{0}^2}{4}} \right),
\end{align}
where we use the Fourier transform of Eq.(\ref{aexp}) in the second line, $\nu \in \{ L,U \}$ indicates the lower and upper bands respectively and $\epsilon_{\nu k}$ is the $U=0$ dispersion shifted by $\mu$. Representing the eigenstates using the momentum-orbital basis

\begin{align}
    b_{\nu k} &\equiv f_{\nu1k} b_{1 k} + f_{\nu2k} b_{2 k}
\end{align}
we find that the lowest order $\tau'$ contribution to the low-energy solutions is

\begin{align}
\label{lowE}
E_{Lk} &\approx 2\tau'' \left( 1 - \gamma^2 \right) \frac{\text{\boldmath$\kappa$}_{\text{slab}}^{2}}{2}\\
\label{lowVUp}
|f_{U1k}|^2 = |f_{L2k}|^2 &\approx \frac{\gamma^2 \text{\boldmath$\kappa$}_{\text{slab}}^2}{2(1+\text{cos}(\kappa_{\perp}))} \rightarrow 0 \\
\label{lowVLow}
|f_{L1k}|^2 = |f_{U2k}|^2 &\approx 1 - \frac{\gamma^2 \text{\boldmath$\kappa$}_{\text{slab}}^2}{2(1+\text{cos}(\kappa_{\perp}))} \rightarrow 1 \\
\label{gamparam}
\gamma &= \frac{\tau'}{\tau''} < 1,
\end{align}
where the diagonal (i.e. $\kappa_{\perp}$-independent) form of the eigenenergies justifies neglecting the lowest order $\tau'$ corrections to the eigenstates and the right hand side of Eqs.(\ref{lowVUp},\ref{lowVLow}) show the low-energy limit of each coefficient. Eqs.(\ref{lowE}-\ref{lowVLow}) conform to our general proofs in the previous sections that the ground states for the $xz$- and $yz$-slabs are completely independent. They further show that the coupling between slabs is suppressed in the low-energy excitations as well.

\subsection{Superfluid Component via Current-Current Response}

\subsubsection{Formulation in Real Space}

We can use the discussion in the last subsection of the first section, on the lack of superfluidity, as a thought experiment to identify the observable corresponding to long-range coherent transport. In the Lehmann representation the \emph{superfluid} component of the response function is given by

\begin{align}
\label{Leh}
& \chi^{\xi \xi'}(\mathbf{r},\mathbf{r'},\omega) = \sum_{\nu k \lambda} \left\{ \frac{\bra{\Omega}j^{\xi}(\mathbf{r})\ket{\nu k \lambda}\bra{\nu k \lambda}j^{\xi'}(\mathbf{r'})\ket{\Omega}}{\omega - (E_{\nu k \lambda} - E_0) + i\eta} - \frac{\bra{\Omega}j^{\xi'}(\mathbf{r'})\ket{\nu k \lambda}\bra{\nu k \lambda}j^{\xi}(\mathbf{r})\ket{\Omega}}{\omega + (E_{\nu k \lambda} - E_0) + i\eta} \right\},
\end{align}
where $\ket{\Omega}$ and $\ket{\nu k \lambda}$ denote the many-body ground state and excited eigenstates of the system respectively, $\mathbf{r},\mathbf{r'}$ are real space lattice points as defined in Eqs.(\ref{Cells1}-\ref{Cells3}), $\xi,\xi'$ are vector components of the current operator. Here $\lambda$ represents a superfluid slab in our model and is simply the 3D superfluid in the case of e.g. a point condensate (where $\sum_{\lambda}$ is then a sum over only one value). The current operator is defined at a lattice site via

\begin{align}
\label{jEq}
    \mathbf{j}_{\alpha XYZ} &= i\sum_{\beta \delta} t_{\beta \delta} \text{\boldmath$\delta$} \left( a^{\dag}_{\beta (XYZ)+\delta} a_{\alpha XYZ} - a^{\dag}_{\alpha XYZ} a_{\beta (XYZ)+\delta} \right).
\end{align}
For example, in our model when orbital $\beta=\alpha=1$ then the matrix element for the superfluid contribution of $\xi$-component of the current reduces to

\begin{align}
    \bra{\Omega}j^{\xi}_{1XYZ}(\tau'')\ket{\nu k \lambda} &= \bra{\Omega} \sum_{\delta \pm \hat{x}} \frac{-i\sqrt{N_Y}\tau''}{\sqrt{L^2}} e^{i\pi X} \Big(\text{\boldmath$\delta$} \cdot \hat{\xi} \Big) \left( a_{1 XYZ} - a_{1 (XYZ)+\delta} \right) b^{\dag}_{\nu k} \ket{\Omega},
\end{align}
where $N_{Y}$ is the reference state density in the slab located at $Y$.

\subsubsection{Confirmation of Standard Superfluid Component}

We first confirm the validity of our formulation using a standard cubic lattice superfluid. In particular, we study the $(\xi,\xi')=(\hat{x},\hat{x})$-component of the current for the one-band cubic lattice with nearest neighbor hopping $t<0$ and repulsive interaction $U>0$. In terms of the eigensolution our formulation gives

\begin{align}
\label{sig1}
    \sigma_{\text{Cube}} (\mathbf{r}) &=  \frac{8t^2 N}{V\sqrt{V}} \sum_{\mathbf{k}} \text{sin}^2(k_{x}) \frac{|c_{\mathbf{k}}|^2}{\Omega_{\mathbf{k}}} e^{-i\mathbf{k}\cdot \mathbf{r}},
\end{align}
where $c_{\mathbf{k}}$ and $\Omega_{\mathbf{k}}$ are a coefficient of the eigenvector and the eigenvalue corresponding to momentum $\mathbf{k}$ respectively and $k_{x}$ is the component of momentum in the $\hat{x}$-direction. This simple one band model is analogous to the low-energy solutions of our two band model in Eqs.(\ref{Hsols}-\ref{Hsolsf}) when $\tau'>\tau''$. We can verify that Eq.(\ref{sig1}) is indeed the correct expression for long-distance coherence by checking the $\mathbf{q}$-component of momentum in the limit $|\mathbf{q}| \rightarrow 0$ to find

\begin{align}
    \lim_{q \rightarrow 0} \sigma_{\text{Cube}} (\mathbf{q}) &= \lim_{q \rightarrow 0} \frac{8t^2 N}{V^2} \sum_{\mathbf{k}\mathbf{r}} \text{sin}^2(k_{x}) \frac{|c_{\mathbf{k}}|^2}{\Omega_{\mathbf{k}}} e^{-i(\mathbf{k}+\mathbf{q})\cdot \mathbf{r}} \\
    &= \lim_{q \rightarrow 0} \frac{8t^2 N}{V} \sum_{\mathbf{k}} \text{sin}^2(k_{x}) \frac{|c_{\mathbf{k}}|^2}{\Omega_{\mathbf{k}}} \left( \frac{1}{V} \sum_{\mathbf{r}} e^{-i(\mathbf{k}+\mathbf{q})\cdot \mathbf{r}} \right) \\
    \label{sqlattfin}
    &= \lim_{q \rightarrow 0} \frac{8t^2 N}{V} \text{sin}^2(q) \frac{|c_{\mathbf{q}}|^2}{\Omega_{\mathbf{q}}} = \frac{2tN}{V},
\end{align}
where we have used the fact that $\text{sin}^2(q)|c_{\mathbf{q}}|^2/\Omega_{\mathbf{q}} \rightarrow 1/4t$ in this limit. Eq.(\ref{sqlattfin}) is the standard result for this model and indicates that all particles are in the superfluid component at $T=0$ and for small enough $U>0$.

\subsubsection{Zero Superfluid Component in our Bose Metal}

Now that we have confirmed that Eq.(\ref{suppcoh}) reproduces the correct superfluid response in real space, we apply it to our model. Since Eqs.(\ref{lowE}-\ref{lowVLow}) show that the eigensolutions are independent of $\tau'$ to first order in our controlled approximation, we can safely neglect $\tau'$ and study the $\tau''$ component of the current operator $\mathbf{j}$ defined in Eq.(\ref{jEq}). Studying the longitudinal response along a diagonal direction (e.g. $(\xi,\xi')=((\hat{y}-\hat{x})/\sqrt{2},(\hat{y}-\hat{x})/\sqrt{2})$ as in Fig.(\ref{suppexpt})) we find

\begin{align}
    \label{sig2}
    \sigma_{\tau''>\tau'} (\mathbf{r}) &= \frac{8\tau''^2 M}{V\sqrt{V}} \sum_{\mathbf{k}} \text{sin}^2 \kappa_{\parallel} \frac{|c_{\text{\boldmath$\kappa$}_{\text{slab}}}|^2}{\Omega_{\text{\boldmath$\kappa$}_{\text{slab}}}} e^{-i\mathbf{k}\cdot \mathbf{r}},
\end{align}
where $M$ is the number of particles in a slab (e.g. $N_{Y}$ and $N_{X}$ for orbital $1$ and $2$ respectively), $\kappa_{\parallel}$ is the component of $\text{\boldmath$\kappa$}_{\text{slab}}$ in the $xy$-plane. Eq.(\ref{sig2}) looks very similar to Eq.(\ref{sig1}), but it is qualitatively different. In Eq.(\ref{sig1}), the reference state is at the $\Gamma$ point in momentum space, so only $\mathbf{k}$ near that point can have low energy excitations. On the other hand, Eq.(\ref{sig2}) comes from eigensolutions that are only functions of $\text{\boldmath$\kappa$}_{\text{slab}}$ and therefore indepenent of $Y$. This freedom implies that solutions along $\mathbf{k}$ are degenerate even at arbitrarily low energy. Once again, this is an expected outcome for a system of effectively independent slabs such as we have here. 

Reorganizing this result to isolate the effect of the independent slabs we find

\begin{align}
\label{sig2fin}
    \sigma_{\tau''>\tau'} (\mathbf{r}) &= \frac{8\tau''^2 M}{V} \left( \frac{1}{\sqrt{2L^2}} \sum_{\text{\boldmath$\kappa$}_{\text{slab}}} \text{sin}^2 \kappa_{\parallel} \frac{|c_{\text{\boldmath$\kappa$}_{\text{slab}}}|^2}{\Omega_{\text{\boldmath$\kappa$}_{\text{slab}}}} e^{-i\text{\boldmath$\kappa$}_{\text{slab}} \cdot \mathbf{r}} \right) \left( \frac{1}{\sqrt{L}} \sum_{k_\perp} e^{-i\kappa_{\perp} r_{\perp}} \right),
\end{align}
where $r_{\perp}$ lies along $\kappa_{\perp}$ as defined in Eq.(\ref{kperp}). Eq.(\ref{sig2fin}) shows that each slab will produce the same result as in Eq.(\ref{sqlattfin}). However, there is an independent sum over $k_{\perp}$ with $r_{\perp}>0$, so the last term vanishes identically. Therefore the superfluid response vanishes between any two different slabs. This tediously derived result is completely consistent with the simpler and more general proof we derived above showing that the slabs decouple at low-energy and therefore cannot host coherent transport in the low-temperature limit.


\end{document}